\documentclass[12pt,preprint]{aastex}
\usepackage{psfig}
\newcommand{\beq}{\begin{equation}}
\newcommand{\eeq}{\end{equation}}
\newcommand{\beqn}{\begin{eqnarray}}
\newcommand{\eeqn}{\end{eqnarray}}

\newcommand{\no}{\noindent}

\newcommand{\sw}{{\rm sw}}
\newcommand{\AU}{{\rm AU}}

\begin{document}
\title{\bf{On Grain Dynamics in Debris Discs:\\ 
Continuous Outward Flows and Embedded Planets} }

\author{Ing-Guey Jiang$^{1}$ and 
     Li-Chin Yeh$^{2}$}

\affil{
{$^{1}$ Department of Physics and Institute of Astronomy,}\\
{ National Tsing-Hua University, Hsin-Chu, Taiwan} \\
{$^{2}$ Department of Applied Mathematics,}\\
{ National Hsinchu University of Education, Hsin-Chu, Taiwan}\\     
}
\authoremail{jiang@phys.nthu.edu.tw}

\begin{abstract}
This study employed grain dynamic models to examine the density distribution 
of debris discs, and discussed the effects of 
the collisional time-intervals of asteroidal bodies,
the maximum grain sizes,
and the chemical compositions of the dust grains of the models, in order to 
find out  whether a steady out-moving flow with an
$1/R$ profile could be formed. The results showed that 
a model with 
new grains every 100 years, a smaller maximum  grain size,
and a composition C400 has the best 
fit to the $1/R$ profile because: (1) the grains have larger
values of $\beta$ on average,therefore, they can be blown out easily;
(2) the new grains are generated frequently enough to replace those 
have been blown out. 
With the above two conditions,
some other models can have a steady out-moving flow with an 
approximate $1/R$ profile. 
However, those models in which new grains are generated every 1000 years
have density distributions far from the profile of 
a continuous out-moving flow. Moreover, the analysis on the signatures
of planets in debris discs showed that there are no indications when a 
planet is in a continuous out-moving flow, however, the 
signatures are obvious in a debris disc with long-lived grains.


\end{abstract}


\newpage
\section{Introduction}

Vega, one of the brightest stars in the Solar neighborhood, has 
became a typical example of stars having discs of dust
due to large infrared excess, as attributed to thermal dust emissions,  
discovered by the Infrared Astronomical Satellite (Aumann et al. 1984).
After that, many other main-sequence stars observed from optical
to submillimeter wavelengths, and revealed dusty disc-like structures, 
thus, were named ``Vega-like stars''.

Debris discs are the dust discs that surround these ``Vega-like stars''.
It is still unclear how these debris discs form.
Naturally, one would expect them to be a product 
of the processes of star formation.
The stars are formed through the collapse of a molecular cloud,
which is a mixture of dust and gas, with a mass ratio about
0.01, as implied by the compositions of the interstellar medium.
The dust grains embedded in the collapsing cloud are the seeds that grow
into larger grains and planetesimals.
In the standard scenario, debris discs are constructed at the time when 
planetesimals are frequently forming and colliding. Thus, 
debris discs can be generated only when there are km-sized planetesimals 
colliding and producing huge amount of new dust grains.
This would take place at the stellar age of a million years,
when the original seed grains grow to become
km-sized planetesimals (Cuzzi et al. 1993).
Moreover, in addition to creating new dust grains, 
the planetesimals would further grow into asteroids and trigger the
formation of planets. In the end, the gaseous parts are 
gradually depleted by 
stellar winds, and the debris discs are constructed.

High-resolution images of some debris discs show the 
presence of asymmetric density structures or clumps, and before 
extra-solar planets were discovered by the Doppler Effect,
these clumpy structures gave indirect evidences of the existence of 
planets. If there were no planets around 
Vega-like stars, it would be much more difficult to explain the 
asymmetric structures of debris discs. 

Astronomers' observational efforts 
have led to  rapid progress on the discovery of planets,
and there are now more than 200 detected 
extra-solar planetary systems. Many theoretical works on their
dynamic structures have been written (Laughlin \& Chambers 2001, 
Kinoshita \& Nakai 2001, Gozdziewski \& Maciejewski 2001, Jiang et al. 2003, 
Ji et al. 2002, Zakamska \& Tremaine 2004, and  Ji et al. 2007). 
The possible effects of discs on the evolution of planetary systems
are also investigated (Jiang \& Yeh 2004a, 2004b, 2004c), and in fact, some of these systems are associated
with the discs of dust. For example,using a sub-millimeter camera, 
Greaves et al. (1998) detected dust emissions around the nearby star
Epsilon Eridani. This ring of dust is at least 0.01 Earth Mass and the peak
is at 60 AU, and it is thus claimed to be a young analog to the Kuiper
Belt in our Solar System. Furthermore, Hatzes et al. (2000) discovered
a planet orbiting Epsilon Eridani by radial velocity measurements,
making the claim by Greaves et al. (1998) even more impressive.

Therefore, the existence of debris disc implies the presence of 
planetesimals, and probably  planets. 
The study of debris discs is very interesting and important
because the density structures and evolutionary histories of debris
discs actually provide hints to the evolution of planetesimals and the
formation of planets.  
Since the Vega system has one of the closest and brightest debris discs, 
many observations have been performed, which reveal  detailed information
(See Harvey et al. 1984, Zuckerman \& Becklin 1993, 
Van der Bliek et al. 1994, Heinrichsen et al. 1998,
Mauron \& Dole 1998, Holland et al. 1998, 
Koerner et al. 2001, Wilner et al. 2002). 
Moreover, Wilner et al. (2002)
showed that the two clumps within  Vega's inner disc could be theoretically
explained by the resonance with a
Jupiter-mass planet in an eccentric orbit. 

In addition to the Vega system, Artymowicz (1997) and
Artymowicz \& Clampin (1997)
discussed the dust discs around $\beta$ Pic, Fomalhaut, and $\alpha$ Lyr.
Grigorieva,and  Artymowicz \&  Thebault (2007) simulated 
collisional dust avalanches of debris discs.
Takeuchi \& Lin (2002) employed a simplified model to study 
the dynamics of dust grains in gaseous proto-stellar discs.
Using a disc model analogous to the primordial solar nebula, they examined the 
effect of a dust grain's size on the dust's radial migrations. 
In principle, the particles at high altitudes move outward, and
the ones at lower altitudes move inward.
In fact, Takeuchi \& Artymowicz (2001) also investigated the same problems.


Interestingly, Su et al. (2005) 
showed the 
images of Vega, as observed by the Spitzer Space Telescope,
and confirmed 
that the size of a Vega debris disc is much larger than previously thought.
Furthermore, from the radial profiles of surface brightness, 
they suggested several models fits, with different combinations of grain 
sizes, and all models require an inverse radial ($1/R$) 
surface number density profile. 
Asteroidal bodies between 86 and 200 AU continue to produce new grains, which  
migrate outward and form an $1/R$ density profile of the outer disc. Most
grains are blown outward, and
their lifetime on the debris disc is relatively short, i.e. less than
1000 years.

However, the above configuration derived from the observations raises a few
important questions on  debris discs in general.  
What is the necessary condition to produce  
the $1/R$ dust density profile ? 
How important are the effects of chemical composition  ?
How does the grain size affect the dust density profile ?
How frequently must  collisions occur in order to produce enough
new dust grains to maintain the $1/R$ profile ?

In order to clarify the above issues,
this study aimed to find a 
self-consistent dynamic models for  debris discs, and, assumed that  
the asteroidal bodies within the inner disc continue generating new grains 
through their collisions.
These grains are then added into the system and move to where it should be
according to the equations of motion. The distributions of these grains
are examined to see whether they follow $1/R$ profiles.
There are many physical processes and parameters to be explored for 
the above models. However, this paper particularly
addresses the effects of collisional time-intervals of asteroidal bodies,
the effects of maximum grain sizes, and the
influences of the chemical compositions of dust grains. 
 
The remainder of this paper is organized as follows. Section 2
presents the model and initial conditions; Section 3 describes the
simulations of a continuous flow; Section 4 discusses
the possible signatures of planets; Section 5 gives the conclusions.

\section{The Model Construction} 

This study aimed to investigate the possible 
self-consistent dynamic models of 
dust distribution on debris discs.
The mass of the dust grains on a debris disc 
could be  $3 \times 10^{-3} M_{\oplus}$ (Su et al. 2005), thus,
if the density of each dust grain is 3.5 ${\rm g/cm^{3}}$ 
and the size is 2 $\mu m$, the total number of grains is
in the order of $10^{35}$.
This estimated number of dust particles 
is too large for any possible numerical simulations. 
Therefore this paper uses 10000 and 30000  
dust grains to represent the outer debris disc in models.
This number 
is large enough to make the spatial resolution of density distributions
sufficiently high for the purpose in this paper. 
However, we do not mean that each particle represents a body consisting of 
a huge number of dust grains.   
In the simulations, each particle only represents one single dust grain.
Because the dust grains do not influence each other in the models,
we could use a 
certain number of them as tracers for the system.
In other words, in our simulations, 
only  density distribution is important, and
the total mass of grains is irrelevant.

This paper focuses on those effects that influence orbital evolution and 
density distributions 
of dust grains. We plan to study ;
(1) the time-intervals between successive collisional events
of asteroid bodies, i.e. the frequency of the generation of new grains;
(2) the effect of maximum grain sizes; and
(3) the influence of chemical compositions.
To complete the above three studies, this paper chooses 
2 chemical compositions (C400 and ${\rm MgFeSiO_4}$),
2 maximum grain sizes
($a_{\rm max}=9.57 \mu m$, and $14.04 \mu m$),
and 2 time-intervals (100 and 1000 years) between successive  
collisions of asteroid bodies
in simulations.

Thus, there will be eight models, and their results
would provide opportunities to connect the grains' orbits and
the density distribution
of debris discs with the three physical ingredients.
In general, whether the grains form a continuous out-moving flow and 
approach a steady $1/R$ density profile will be examined. 
For convenience, a simulation with C400, a time interval of ${10}^2$
years, and a smaller value of maximum gain sizes are presented in 
Model C2S. Similarly, Model Mg3L stands for a simulation
with  ${\rm MgFeSiO_4}$, a time interval of ${10}^3$
years, and a larger value of maximum gain sizes. 
Table 1 lists these eight models.

In our models, the dust grains' motion is governed by gravity and radiation
pressure from the central star. 
Through the calculations of the orbital evolution of these dust grains,
the density distribution of the debris disc at any particular time
could be determined. 

\subsection{The Units}

For the equations of motion, the unit of mass is $M_{\odot}$,  
the unit of length is AU, and the unit of time is a year. 
Thus, the gravitational constant
$G=6.672\times 10^{-11} ({\rm m^3/kg\,sec^2})
=38.925({\rm (AU)^3/M_{\sun} year^2})$, and the light speed 
$c=3\times 10^8{\rm  (m/sec)}=6.3\times 10^4 ({\rm AU/year})$.

\subsection{The Equations of Motion}

All dust grains are assumed to be in a 
two dimensional plane, governed by  gravity and radiation pressure
from the central star. 
For any given time, the central star is fixed at the origin, and 
the dust grain's equations of motion are, 
as in Moro-Martin \& Malhotra (2002):

\beq\left\{
\begin{array}{ll}
&\frac{d^2 x}{dt^2}=-\frac{Gm_0(1-\beta)}{R^3}x-\frac{\beta_\sw}{c}
\frac{Gm_0}{R^2}\left[\left(\frac{\dot{R}}{R}\right)x+\frac{dx}{dt}\right],\\
&\frac{d^2 y}{dt^2}=-\frac{Gm_0(1-\beta)}{R^3}y-\frac{\beta_\sw}{c}
\frac{Gm_0}{R^2}\left[\left(\frac{\dot{R}}{R}\right)y+\frac{dy}{dt}\right],\\
\end{array}\right.\label{eq:ini1}
\eeq
where,
\beq\left\{
\begin{array}{ll}
&R=\sqrt{x^2+y^2}, \\
&\dot{R}=\frac{x}{R}\frac{dx}{dt}+\frac{y}{R}\frac{dy}{dt},\\
&\beta_\sw=(1+\sw)\beta,
\end{array}\right.\label{eq:dot_r}
\eeq 
further, $(x,y)$ is the coordinate of a particular grain, 
$G$ is the gravitational constant, $m_0$ is the central star's mass
(2.5$M_{\sun}$), 
$c$ is the speed of light, $\beta$ is the ratio between the 
radiation pressure force and the gravitational force,
and ``sw'' is the ratio of the solar wind drag 
to the P-R drag. In this paper, ${\rm sw}$ is taken to be zero.

\subsection{The Optical Parameters of Dust Grains} 

The equations of motion 
show that, when a parameter is $\beta < 0.5$, the grain's orbit 
is likely to be bounded; when $\beta > 0.5$, the grain would have an 
unbound orbit.
To demonstrate the effect of $\beta$ on the orbital evolution, the 
evolution of radial 
velocities and distances are presented in Fig. 1.  
Fig. 1(a) shows
the radial velocities of grains with given $\beta$ as functions of time
when they are initially located at $R=100$ AU. 
For a grain with $\beta=0.35$, the radial velocity
can be positive or negative, and oscillate around zero. It moves on 
an elliptical orbit because it has a negative total energy.
A grain with $\beta=0.45$ has similar behavior, however, the orbital period
is much longer. For a grain with $\beta=0.55$, the orbit becomes unbound
and the radial velocity approaches a constant value, i.e. terminal velocity.
Similarly, grains with $\beta=0.65, 1.0, 1.5, 2.0$, and $2.5$ also 
have unbound orbits and approach terminal velocities. 
The curves in Fig. 1(a) show that 
grains with larger $\beta$ have larger terminal velocities.
Figs. 1(b), 1(c), and 1(d) show the radial distances of grains 
with given $\beta$ as functions of time
when they are initially located at $R=100$, 200, and 300 AU.
The solid curves are for $\beta=0.2$, dotted curves are for
$\beta=0.3$, dashed curves are for $\beta=0.4$, and the long dashed
curves are for $\beta=0.5$. 
The grains with $\beta=0.5$ all escape from the systems.
For the grains with $\beta=0.4$, they reach
 500 AU if initially located at 100 AU, reach  1000 AU if initially set at
200 AU, and reach 1500 AU when initially put at  300 AU.

The parameter $\beta$ in the equations of motion determines how important 
the radiation pressure is and can be calculated as (Burns et al. 1979):
\beq
\beta= \frac{3L}{16\pi G m_0 c\rho a} 
\frac{\int Q_{pr}(a,\lambda) F_{\lambda} d\lambda}
{\int F_{\lambda} d\lambda},
\eeq
where, $G$, $m_0$ and $c$ are as previously defined.
$L$ is the central star's luminosity, $\rho$ is the dust grain's density,
$a$ is the radius of the dust particle, $F_{\lambda}$ is the central star's
spectrum. The radiation pressure factor of optical parameters, 
$Q_{pr}(a, \lambda)$, is a function of grain's radius $a$ and the incident 
electromagnetic wave's wavelength $\lambda$, and  depends on the
chemical composition of the considered grain. In general, the smaller grains
would have larger $\beta$.
In this paper, we choose C400 and ${\rm MgFeSiO_4}$ as the compositions
of the dust grains to calculate their corresponding $\beta$. The reason why
we choose these two is that C400's values of $\beta$ are among one of the 
largest, and ${\rm MgFeSiO_4}$'s values of $\beta$ are around one of 
the smallest (as seen in Fig. 5 of Moro-Martin et al. 2005). 
According to Laor \& Draine (1993), 
the density of C400 grains is 2.26 ${\rm g/cm^3}$, and the one of 
${\rm MgFeSiO_4}$ grains is 3.3 ${\rm g/cm^3}$. 
Then, Mie Scattering Theory and Vega's spectrum
are used to determine $\beta$ of a particular grain
(please see Appendix A for details).

\subsection{The Grain Size Distributions} 
  
As shown in Fig. 1, the non-gravitational 
influence on the grains' orbital evolution is completely determined 
by the parameter $\beta$. When the chemical compositions of dust grains
are chosen, the values of $\beta$ mainly 
depend on the grain sizes. 
For the size distributions in this study, the classic standard 
power law, with an index -3.5 is applied, i.e : 
\beq
\frac{dN}{da} = C a^{-3.5},
\eeq
where $N$ is the grain number, $a$ is the grain radius, $C$ is a constant.
Numerically, it can be written as :
\beq
\triangle N = C a^{-3.5} \triangle a, \label{eq:gsize1}
\eeq
where $\triangle a$ is the chosen bin size and $\triangle N$
is the expected grain number in the bin with a grain size around $a$.


We set $a_{\rm bottom}=1 \mu m$ and $a_{\rm up}=46 \mu m$ and 
choose a uniform bin size in the logarithmic space as
$(\ln a_{\rm up}-\ln a_{\rm bottom})/100=0.038286$. 
We then have  
$a_i=\exp\{0.038286\times (i-1)\}$ for $i=1,2,\cdots, 100$ and define 
$\bar{a_i}=(a_{i+1}+a_{i})/2$ for $i=1,2,\cdots, 99$ 
as the possible grain sizes.
By  Eq.(\ref{eq:gsize1}), 
 $$\triangle N_i(\bar{a_i})=C\bar{a_i}^{-3.5}(a_{i+1}-a_{i}) 
\quad {\rm for}\  i= 1,2,\cdots,99.$$
From the above, we have
\beq
N_{\rm tot}\equiv\sum_{i=1}^{i=99}\triangle N_i(\bar{a_i})=C \sum_{i=1}^{i=99}
\bar{a_i}^{-3.5}(a_{i+1}-a_{i}).  \label{eq:c}
\eeq
Once the total particle number $N_{\rm tot}$ is given, the parameter $C$
can be determined from Eq.(\ref{eq:c}). 
Thus,
we set the first grain size as $\bar{a_1}$, and 
the number of this size of grains to be  
${\rm INT}[\triangle N_1(\bar{a_1})]+1
={\rm INT}[C\bar{a_1}^{-3.5}(a_2-a_1)]+1$,
where INT is an operator to take the integer part of a real number.
The 2nd grain size is $\bar{a_2}$ and the number of this size is similarly 
determined. We continue this process until the total number of grains 
approaches $N_{\rm tot}$ as possible as it can be.
For example,  
in this paper, when $N_{\rm tot}=10000$, 
we start from the first grain size $\bar{a_1}$
until $\bar{a_{59}}$. We find $\triangle N_{59}(\bar{a_{59}})=3.54$,
so the number of grains with size  $\bar{a_{59}}$ is 4. At this stage,
the total number of grains is 9996.
Luckily, $\triangle N_{60}(\bar{a_{60}})=3.21$, therefore, we set the number 
of grains with size $\bar{a_{60}}$ as 4 and the total number of grains
is 10000. Please note that $\bar{a_{60}}= 9.57 \mu m$, which is smaller
than $a_{\rm up}$. Thus, in our simulations, when 
the total grain number is 10000, the maximum
grain size $a_{\rm max}$ is $9.57 \mu m$.
When $N_{\rm tot}=30000$,  we proceed similarly and 
 find $\triangle N_{69}(\bar{a_{69}})=4.08$.
The number of grains with size $\bar{a_{69}}$ is  5 and the total number is
now 29997. Although  $\triangle N_{70}(\bar{a_{70}})=3.7$, we still
set the number of grains with size $\bar{a_{70}}$  to be 3 only, 
in order to make
the total number of grains be 30000.
Thus, the maximum grain size $a_{\rm max}$ is $\bar{a_{70}} = 14.04 \mu m$
when  $N_{\rm tot}=30000$.



Fig. 2(a-1) shows the number of grains as a function of grain size
of models with $a_{\rm max}=9.57 \mu m$ 
and Fig. 2(b-1) shows the one of models with $a_{\rm max}=14.04 \mu m$.
Fig. 2(a-2) and 2(b-2) show the histograms of the grains' 
corresponding $\beta$ values.
One can see that 
the $\beta$ values of C400 grains, triangles, are larger
than the ones of ${\rm MgFeSiO_4}$ grains, circles,
in both Fig. 2(a-2) and 2(b-2). 

\subsection{The Initial Distributions}

The dust particles are supposed as produced through the collisions of 
asteroidal bodies in the ring region, between 86 and 200 AU.
The initial positions of  
the dust grains in all models are therefore placed in this region.  
From 86 to 100 AU, the surface number density is set to be a constant.
At $R=100\AU$ the surface number density starts to 
decrease as $1/R^2$ until $200\AU$. Thus, the surface number density is :
\beq
\Sigma_{1}(R)=\left\{
\begin{array}{lll}
 &\frac{\Sigma_0}{100^2}   &{ \rm when \,\, }86 \le R \le 100,\\
& \frac{\Sigma_0}{R^2} &  {\rm when\,\, }  100< R \le 200,
\end{array}\right. \label{eq:sigma1}
\eeq
where $\Sigma_0$ is a constant.
At the beginning of the simulations, i.e. $t=0$, 
the initial particle number is $N_{ini}=10000$
for those models with $a_{\rm max}=9.57 \mu m$,  
and $N_{ini}=30000$ for models with 
$a_{\rm max}=14.04 \mu m$.
Thus, the constant  $\Sigma_0$ can be determined by :
\beq
N_{ini}=\int_{86}^{100}2\pi R\frac{\Sigma_0}
{(100)^2} dR +\int_{100}^{200}2\pi \frac{\Sigma_0}{R}dR. 
\eeq

After that, we add $N_{ini}$ grains into the system 
with the above distribution 
at $t=100 \times i$, $i=1, 2,..., 100$ (for Model 
C2S, C2L, Mg2S, Mg2L) or 
at $t=1000 \times i$, $i=1, 2,..., 10$ (for Model 
C3S, C3L, Mg3S, Mg3L). 
Thus, the total simulation time would be $10000$ years.




The basic ingredients of all models are summarized in Table 1.\\

\centerline{ {\bf Table 1} The Ingredients of Models} 
\begin{center}
\begin{tabular}{|c|c|c|c|c|c|}
\hline
Model & Composition & Grain Density&  Time Interval
& $a_{\rm max}$ & $\beta_{\rm min}$     \\\hline
C2S & C400        & 2.26(${\rm g/cm^3}$)  &  
100 (years)  &  9.57 ($\mu m$)   & 0.62   \\ \hline
C2L & C400        & 2.26(${\rm g/cm^3}$)  &  
100 (years)  &  14.04 ($\mu m$)        & 0.42   \\ \hline
C3S & C400        & 2.26(${\rm g/cm^3}$)  &  
1000 (years)  &  9.57 ($\mu m$) & 0.62   \\ \hline
C3L & C400        & 2.26(${\rm g/cm^3}$)  &  
1000 (years)  &  14.04 ($\mu m$) & 0.42   \\ \hline
Mg2S &${\rm MgFeSiO_4}$  & 3.3(${\rm g/cm^3}$) &  
100 (years)  & 9.57 ($\mu m$) & 0.43      \\ \hline
Mg2L &${\rm MgFeSiO_4}$  & 3.3(${\rm g/cm^3}$) &  
100 (years)  &  14.04 ($\mu m$) & 0.29      \\ \hline
Mg3S &${\rm MgFeSiO_4}$  & 3.3(${\rm g/cm^3}$) &  
1000 (years)  & 9.57 ($\mu m$) & 0.43      \\ \hline
Mg3L &${\rm MgFeSiO_4}$  & 3.3(${\rm g/cm^3}$) &  
1000 (years)  &  14.04 ($\mu m$)  & 0.29      \\ \hline
\end{tabular} \\
\end{center}

\subsection{The Initial Velocities}

The asteroidal bodies in the ring region could move on any orbits, but their 
average velocities should be 
close to the velocities of circular motions.
To simplify the models, we assume all dust grains, which are supposed as 
generated from the larger asteroidal bodies, move on circular orbits
initially.

\subsection{Fitting Functions and Scaling Factors}

In order to determine  whether the surface mass density of the 
simulation result follows $1/R$ distributions, 
we must first determine the
best $1/R$ fitting functions.
Following the least square method of Cheney and Kincaid (1998), 
the function $\phi$ is defined by :  
\beq
\phi(c)=\sum^{n}_{i=1}\left\{\frac{c}{R_i}-S_i\right\}^2
\eeq
where $S_i$ is the surface mass density at the radius $R_i$
(The disc is now separated into $n$ annulus.) and
the surface mass density is considered beyond 200 AU for this fitting. 
The best value of $c$ can be determined through 
$d\phi/dc=0$, and this value $c_s$ divided by $R$
is the best $1/R$ fitting function, $c_{s}/R$.

\subsection{Low Collisional Probability between Dust Grains}

The estimation of possible collision rates between dust grains
could be complicated as it is related to the grain sizes, the spatial
distributions, and the total mass. 
Thebault and Augereau (2007) conducted simulations to address this important
issue, and defined a collisional lifetime $t_{\rm coll}$, 
which is the average time it takes
for an object to lose 100 $\%$ of its mass by collisions.
The results of the collisional lifetimes are presented in
their Fig. 4, where a debris disc with 
a total mass $0.1 M_{\oplus}$ (left panel) and $0.001 M_{\oplus}$
(right panel) are both considered.
The total mass of the debris disc considered in this paper 
is about $3 \times 10^{-3} M_{\oplus}$, therefore, the right panel of their
Fig. 4  fits our case.
Because the grain size in our models is around $10 \mu m$, 
the $t_{\rm coll}$ is from $2 \times 10^4$ years to $2 \times 10^6$ years.
As the simulation time of our models will be only 4000 years,
the collisional lifetime $t_{\rm coll}$ is much longer than the timescale
we consider in our simulations.
Therefore, the possible collisions between dust grains
are ignored in this paper.

\section{Simulations of a Continuous Out-Moving Flow}

This section will discuss 
the grains' distribution on the disc for all simulation models. 
The simulations started at $t=t_0\equiv 0$, and terminated at 
$t=t_{\rm end}\equiv 4000$ years.
To clearly present the evolution during the simulation,
the dust distribution is plotted every 200 years for 20 panels. 
 
In order to understand the evolution of both long-lived and 
short-lived grains,  the 
grains are divided into two groups according to their $\beta$, i.e. 
the larger grains are those with $\beta < 0.5$ and  
the smaller grains are those with $\beta \ge 0.5$.
Table 2 gives the number percentages of the  
smaller and larger grains in our models.\\

\centerline{ {\bf Table 2} Number Percentages of Grains } 
\begin{center}
\begin{tabular}{|c|c|c|}
\hline 
Model & smaller grains ($\beta \ge 0.5$) & larger grains ($\beta<0.5$) 
 \\ \hline
C2S, C3S   & 100\%   & 0\%    \\ \hline
C2L, C3L   & 99.93\% & 0.07\% \\ \hline
Mg2S, Mg3S & 99.03\% & 0.97\% \\ \hline
Mg2L, Mg3L & 98.88\% &  1.12\%  \\ \hline
\end{tabular}
\end{center}
Accordingly, in all
the plots of grains' density distributions, 
the crosses are for the larger grains, the triangles
are for the smaller grains, and the circles are the total surface mass density.


Fig. 3 shows the surface mass density of dust grains distributed
between 200 and 1400 AU at $t=200\times i$ years, where
$i=1, 2, 3,..., 20$, for Model C2S. 
Since all grains are located
between 86 and 200 AU initially, there is no grain in the region beyond 200 AU
at $t=0$. To save the space, 
the grain distribution at time $t=0$ is not plotted. 
As shown in  Table 2, there are no larger grains 
(i.e. those with $\beta<0.5$) 
in this model, so the crosses are always at the value of zero. 
Panel 1 shows that at $t=200$, the smaller grains spread over the region 
between 200 and 500 AU.
Since new grains are added into the system every 100 years, 
even more grains appear
in the region between 200 and 500 AU and some grains migrate even furthermore,
as shown in Panel 2, 3, and 4. While more and more grains move outward, 
the whole outer disc beyond 200 AU can be well fitted by an $1/R$ profile
by time $t=2000$ (Panel 10), and after, the outer disc becomes
a steady continuous flow and always follows an $1/R$ profile.

In order to understand disc evolution
when collisions between asteroid bodies occur much less often, we also 
do simulations as new grains are added every 1000 years. Model C3S
is for this purpose, and 
the results are presented in Fig. 4. 
At $t=200$, as shown in Panel 1,
some initial grains move to the region between 200 and 500 AU. 
These grains move outward even further in the following panels, however, 
the surface
mass density decays as there are no new grains to maintain the
profile between 200 and 500 AU. At $t=1000$ (Panel 5), 
the density profile decays
to nearly flat, and 10000 new grains are added at that time.
These new grains migrate outward, thus, the surface mass 
density between 200 and 500 AU raises in Panel 6. 
In general, Panels 6 to 10 repeat the decaying seen in Panels 1 to 5.
Similar processes happen from Panels 11 to 15, and Panels 16 to 20.
The $1/R$ profile does not have good fitting function for the surface mass 
density at any time in this model. 

Fig. 5 shows the evolution of the disc's surface mass density of Model C2L,
which is a model with a small fraction of larger grains ($0.07 \%$)
and adding new grains every 100 years.
At $t=200$, some smaller grains migrate up to about 500 AU
but there is almost no larger grains appear in this region.
At $t=400$ (Panel 2), the surface mass density of smaller grains increases
and some larger grains migrate up to around 200 and 300 AU.
Both the smaller and larger grains continue moving outward, and the overall 
surface mass density approaches 
a decaying function with some fluctuations. 
Because most of the larger grains
are bounded within the system, only the smaller grains form a steady 
out-moving flow.
The existence of larger grains makes it more difficult for the total surface 
mass density  to become an $1/R$ profile. 
It cannot be fitted by an $1/R$ function until $t=3200$ (Panel 16),
and density fluctuation deviation from the $1/R$ curve 
is larger than  in Model C2S.

Fig. 6 presents the evolution of the grain's surface mass density of Model C3L.
As in Model C3S, the profile cannot be maintained 
due to new grains not being added with enough frequency.
The main difference is that the larger grains exist and stay in the 
region between 200 and
600 AU. Their persistence makes the total surface  mass density 
closer to the $1/R$ profile at some particular time. 
For example, the distribution in Panel 7 and 12 are closer to the $1/R$ 
profile, however, there is no steady out-going flow in this model.

In order to investigate the effects of chemical compositions, 
this paper offers another set of four models Mg2S, Mg3S, Mg2L, and Mg3L
(please see details in Table 1 and Table 2), and their 
results are shown in Figs. 7-10.
Table 2 shows that the fractions of long-lived grains of these four models 
(i.e. those with $\beta < 0.5$)
are larger than those models with C400. 
In general, these larger grains would remain around the system for a time scale
much longer than the smaller grains, and their persistence 
cause the density deviations from the $1/R$ profile. 
For example, Fig. 7 shows that, in Model Mg2S, the smaller grains form a 
continuous out-moving flow starting from $t=3000$ (Panel 15), 
however, the larger grains continue moving out slowly.
The overall density distribution still approaches an $1/R$ profile,
though much slower and with larger fluctuations.
Fig. 8 shows that, in Model Mg3S, the new grains are not generated frequently
enough to form a steady density profile. The surface mass density
is often very small in all areas of the disc.    
Fig. 9 presents the grains' distributions of Model Mg2L, which 
is a model similar to Model Mg2S, but with a greater fraction 
of larger grains.
The contribution of the persistent larger grains make it very difficult
to have an $1/R$ density profile, though the profile approximately
becomes a steady-state after $t=2400$ (Panel 12).
Finally, the results of Model Mg3L are shown in Fig. 10, because  new grains
are added every 1000 years, the density goes up and down randomly.
There are some larger grains remaining, however, the density cannot be 
fitted by an $1/R$ profile.

\section{Signatures of Planets}

As discussed in Section 1, planets could exist in debris discs.  
For example, two clumps of the Vega's inner disc, 
as studied by Wilner et al. (2002), could be due to 
the resonant capture of dust grains by a Jupiter-mass planet.
Thus, the non-axisymmetric
structures of debris discs give obvious signatures of planets
and the resonant trapping is a natural
explanation for these clumpy structures.

In addition to the resonant capture, gravitational scattering
by the planet  also influences  grain distribution of debris discs. 
It is therefore interesting to see what could happen if there is a planet
moving around the debris disc, with the continuous out-moving grain flow 
produced in the previous section. 
 

First, this study examined the disc's structure
in one of the previous models, Model C2S, in the case 
when a five-Jupiter-mass planet is added.
In Run 1, the planet is initially located at $(x,y)=(100 {\rm AU}, 0)$,
and is following  simple circular motions. However, in Run 2, 
the planet is initially located at $(x,y)=(250 {\rm AU}, 0)$, and
the remaining details of the above two runs
are the same as in Model C2S.

Fig. 11 shows the disc's surface mass density 
in Model C2S, Run 1, and Run 2
from $R=0$ to $R=1000$ AU. The circles represent Model C2S, the crosses
represent Run 1, and the triangles represent Run 2.
It is clear that the
discs' profiles in these three models are the same  
at any point during the simulations.
When the planet orbits at $R=100$ AU, as in Run 1, it would not greatly 
affect the grains, as almost all grains leave their birth places 
(86 to 200 AU).
The density peaks were around 250 AU, therefore, we choose the planet to 
move in 
a circular orbit, with a radius of $R=250$ AU, in Run 2.
However, it is found 
that the planet does not affect the disc's density profile.
 

In order to observe the disc's evolution from another view, 
the distribution of dust grains on the $x-y$ plane 
in Run 2 is presented in Fig. 12. The small dots represent the dust grains, 
and 
the full circle represents the planet.
It is obvious that the grain distribution in 
every panel looks almost identical, and
there is no signature for the planet. The corresponding plots
for Model C2S and Run 1 are the same as Fig. 12, and thus, are not 
 shown here.
The planet is hidden among the debris disc with 
a continuous out-moving flow.
This is due to  
the scattering probability 
between the planet and dust grains, which is very small in a continuous 
out-moving flow. The dust grains pass  
by and move outward quickly in a short time. 

Secondly, in order to demonstrate the outcome when  
 the dust grains have a much longer
life-time, in Run 3, a simulation was performed, where 10000 
dust grains were placed in the region between 86 and 200 AU, following
Eq. (7), as in Model C2S.
However, all  dust
grains had $\beta=0.1$, which corresponds to  grain 
radius $a=58.53 \mu m$ for C400, and $a=40.62 \mu m$
for ${\rm MgFeSiO_4}$.
These represent larger dust grains in the ring region, which 
would only have tiny migrations and not be blown out.
In Run 3, a planet with five-Jupiter-mass is initially located at 
$(x,y)=(100 {\rm AU}, 0)$, and move in a circular orbit.

The evolution of grain distributions on the $x-y$ plane for Run 3
is shown in Fig. 13.
The locations of dust grains are shown by dots, and the
full circle represents the planet.
The 1st panel, which is the situation at $t=200$,
shows that dust grains are remaining in their birthplaces.
However, in Panel 2, some grains are scattered by the planet
and move on elongated orbits. In Panels 3 and 4,
the spiral-shape gaps are formed gradually.
Later, a ring-like structure becomes obvious in the 5th and 6th panels.
In Panels 7, 8, and 9, the ring becomes larger
and another arc-like structure forms inside the ring.
Due to the influence of the planet,
these non-axisymmetric structures are developed continuously 
until the end of the simulation.

For purposes of comparison, another simulation was performed as Run 4, 
which is the same as Run 3, except there is no planet in this run.
The evolution of dust grains on the $x-y$ plane for this simulation
is shown in Fig. 14.
Due to  the dust grains being subjected to radiation pressure, 
the orbit expands  slightly, and
the radial oscillations cause rings to form at different places.

To carefully examine particle distribution, the color contours
of the final panel of Fig. 12-14 (i.e. Run 2, 3, 4) 
are shown in top-right, bottom-left, 
and bottom-right of the Figs. 15. In addition, the color contour of 
the final grain distribution of Model C2S is  shown in 
the top-left panel of Fig. 15, for comparison.
Thus, those models with a continuous out-moving flows could have 
a larger disc with a density peak at $R=250$ AU. Their contours look
the same no matter whether there is a planet or not.
The models with long-lived grains have smaller discs,
and the one with a planet added could have non-axisymmetric structures. 



To summarize, from the first two panels of Run 3, it is found that, to make
the planet-grain scattering strong enough to produce 
signatures in debris discs, 
the dust grains have to stay in a close-by orbit for at least 400 years.
When dust grains of the debris disc are short-lived,
they do not stay in the same orbital radius for many periods, and simply
keep moving, and continuously approach 1000 AU.
This is the reason 
there is no signature in a debris disc, 
when a planet is added in an out-moving flow, and it looks as though the planet
is {\it hidden} in a continuous flow.
On the other hand, 
when dust grains are long-lived, the  
non-axisymmetric structures are easily formed due to the scattering 
between the 
planet and dust grains. 










\section{Concluding Remarks}

This study investigated the possibilities of constructing
general self-consistent dynamic 
models of debris discs, and particularly, examined the
effects of collisional time-intervals of asteroidal bodies,
the effects of maximum grain sizes,
and  the influence of chemical compositions of dust grains. 
In all the simulations,
the grains' orbits are calculated, the density distributions
of debris discs are then determined, and compared with 
the $1/R$ functions.

The results showed the grains in the Model C2S give the best 
fit to the $1/R$ profile because : (1) the grains have larger
values of $\beta$ in average, and thus, they can be blown out easily; and
(2) the new grains are generated frequently enough to replace those 
have been blown out.
The above two conditions make it easier to form a continuous out-moving
flow and thus approach an $1/R$ profile. It is worth noting that 
it only takes about 2000 years to become  an $1/R$ profile.
When there are not enough new grains, the profile cannot be 
maintained, as shown in Model C3S and Model C3L.  
The persistence of some larger grains in Model C2L make the density 
profile slightly more complicated.
Those models with ${\rm MgFeSiO_4}$ have more bound grains, therefore,
the deviations from the $1/R$ profile are larger than  with
C400.  To conclude, 
those models in which new grains are generated every 1000 years
have density distributions far from the profile of 
a continuous out-moving flow.

In the study of the signatures
of planets in debris discs, the results showed that there is no 
sign at all when the planet is in a continuous out-moving flow, however, the 
signatures are obvious in a debris disc with long-lived grains.



\section*{Acknowledgment}
We thank the anonymous referee for useful remarks and suggestions that
improved the paper enormously.  
We also owe a debt of thanks to 
A. Moro-Martin, K. Su, and M. Holman, whose communications and conversations
were really helpful.  
We are grateful to the National Center for High-performance Computing
for computer time and facilities.
This work is supported in part 
by the National Science Council, Taiwan, under 
NSC 97-2112-M-007-005.

\section*{Appendix A: Mie Scattering Theory}

The exact solution of the scattering of an electromagnetic wave by a
dielectric sphere of arbitrary size is referred as the
Mie Scattering Theory. 
Using the radial components of the electric and magnetic Hertz vectors 
($\pi_1$, $\pi_2$),
the Hertz potential can be found and the general expressions for the
scattered fields can be obtained (Ishimaru 1991). Applying the boundary 
conditions on $\pi_1$ and $\pi_2$, the constant 
$a_n$ and $b_n$ in the general expressions can be determined.
It is found that both $a_n$ and $b_n$ are related with 
spherical Bessel functions $j_n(x)$ and $h_n^{(2)}(x)$.
The independent variable $x$ is usually called
the size parameter and can be defined as
\beq
x=\frac{2\pi a}{\lambda},
\eeq
where $\lambda$ is
the wavelength of the incident wave, and $a$ is the radius
of the dust grain.
We also need to define $\psi_n(x)$ and $\zeta_n(x)$ as
\beq
\psi_n(x) = x j_n(x)
\eeq
\beq
\zeta_n(x) = x  h_n^{(2)}.
\eeq
The complex index of refraction is $m$, and $y$ is defined as
\beq
y=m x. 
\eeq
Then, we can explicitly express $a_n$ and $b_n$ as
\beq
a_n= 
\frac{\psi_n^{\prime}(y)\psi_n(x)- m\psi_n(y)\psi_n^{\prime}(x)}
     {\psi_n^{\prime}(y)\zeta_n(x)-m\psi_n(y)\zeta_n^{\prime}(x)}\\
b_n=\frac{m\psi_n^{\prime}(y)\psi_n(x)-\psi_n(y)\psi_n^{\prime}(x)}
     {m\psi_n^{\prime}(y)\zeta_n(x)-\psi_n(y)\zeta_n^{\prime}(x)}
\eeq

The extinction factor $Q_{\rm ext}$, 
defined as the total cross section 
($\sigma_t$) divided by the geometric cross section ($\pi a^2$),
can be calculated as (Ishimaru 1991) 
\beq
Q_{\rm ext}= \frac{2}{x^2} {\rm Re} [\sum_{n=1}^{\infty}(2n+1)(a_n+b_n)],
\eeq
where Re means taking the real part only. 
Similarly, the scattering factor $Q_{\rm sca}$, 
defined as the scattering cross section 
($\sigma_s$) divided by the geometric cross section ($\pi a^2$),
can be calculated as 
\beq
Q_{\rm sca}= \frac{2}{x^2} \sum_{n=1}^{\infty}(2n+1)(|a_n|^2+|b_n|^2). 
\eeq
The absorption factor $Q_{\rm abs}$ is defined as 
\beq
Q_{\rm abs} = Q_{\rm ext} - Q_{\rm sca}.
\eeq
Please note that Van De Hulst (1957) also derived the above results similarly
and further define the radiation pressure factor $Q_{\rm pr}$ as 
\beq
Q_{\rm pr} = Q_{\rm ext} 
-\frac{4}{x^2}\sum_{n=1}^{\infty} \frac{n(n+2)}{n+1} 
{\rm Re}(a_n a_{n+1}^{\star} + b_n b_{n+1}^{\star})
-\frac{4}{x^2}\sum_{n=1}^{\infty} \frac{2n+1}{n(n+1)} 
{\rm Re}(a_n b_n^{\star}).
\eeq

The relation between a wavelength and 
the complex index of refraction $m$ for a grain with
a particular chemical composition can be obtained 
from the website, http://www.astro.uni-jena.de/Laboratory/Database/.
We get that for C400 and ${\rm MgFeSiO_4}$ and 
calculate the $Q_{\rm pr}$ and $Q_{\rm abs}$
numerically through the above equations.
The spectrum of Vega is taken from the site,
ftp://ftp.stsci.edu/cdbs/cdbs2/grid/k93models/standards/.

\clearpage
\begin{figure}[htbp]
\psfig{figure=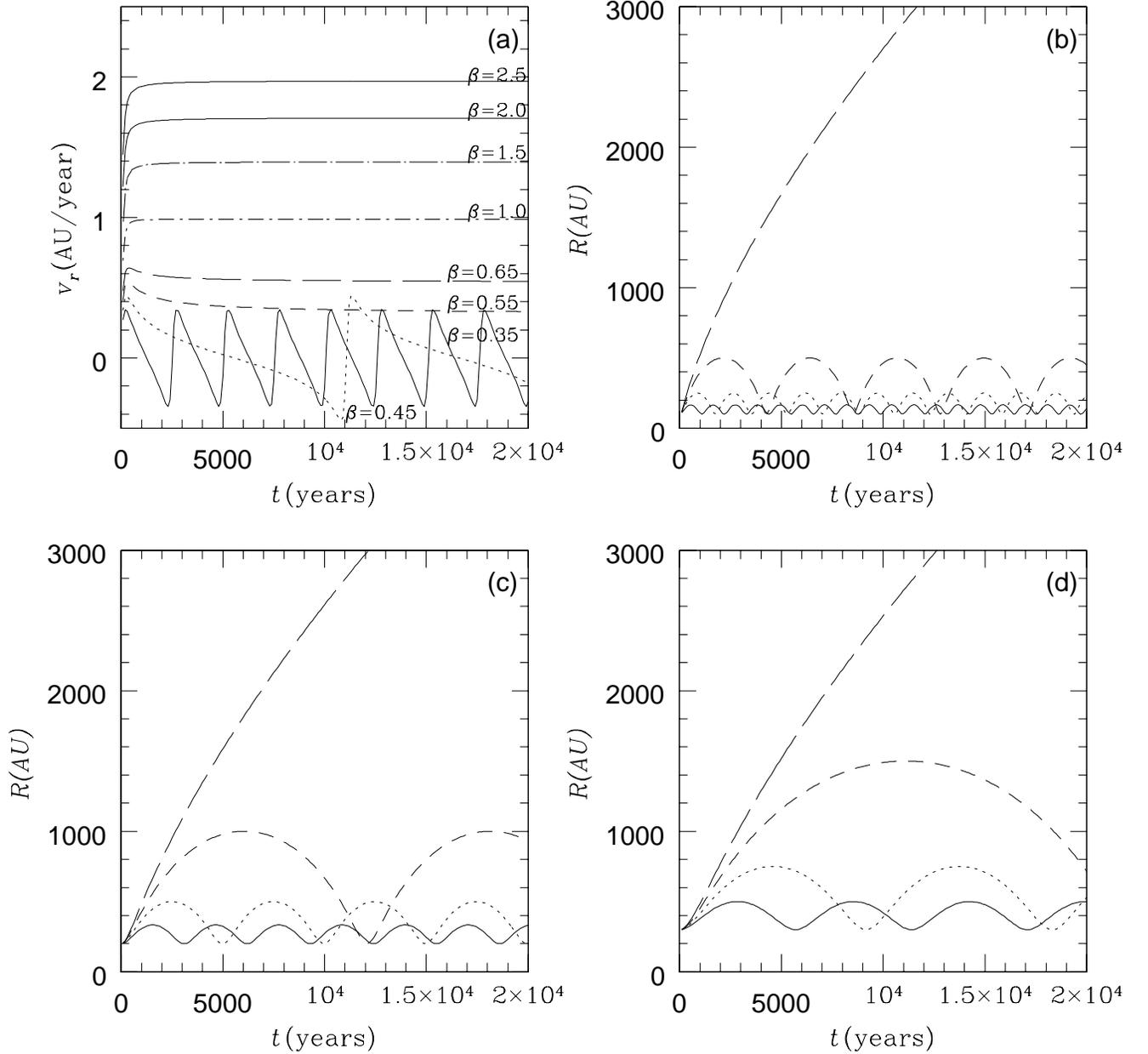,width=18.cm}
\caption[]{(a) The radial velocities, $v_r$,
as functions of time for grains with
different values of $\beta$. The grains are at 100 AU initially.
The bottom solid curve is for $\beta=0.35$.
The dotted curve is for $\beta=0.45$.
The short dashed curve is for $\beta=0.55$.
The long dashed curve is for $\beta=0.65$.
The short dashed-dotted curve is for $\beta=1.0$.
The long dashed-dotted curve is for $\beta=1.5$.
The top two solid curves are for $\beta=2.0$ and $\beta=2.5$;
(b)-(d) The radial distances, $R$, as functions of time for grains with
different values of $\beta$. The grains are initially 
at 100, 200, and 300 AU
for Panel (b), (c), and (d). The solid curves are for $\beta=0.2$,
doted curves are for $\beta=0.3$, dashed curves are for $\beta=0.4$, 
and long dashed curves are for $\beta=0.5$.
}
\end{figure}
\clearpage
\begin{figure}[htbp]
\psfig{figure=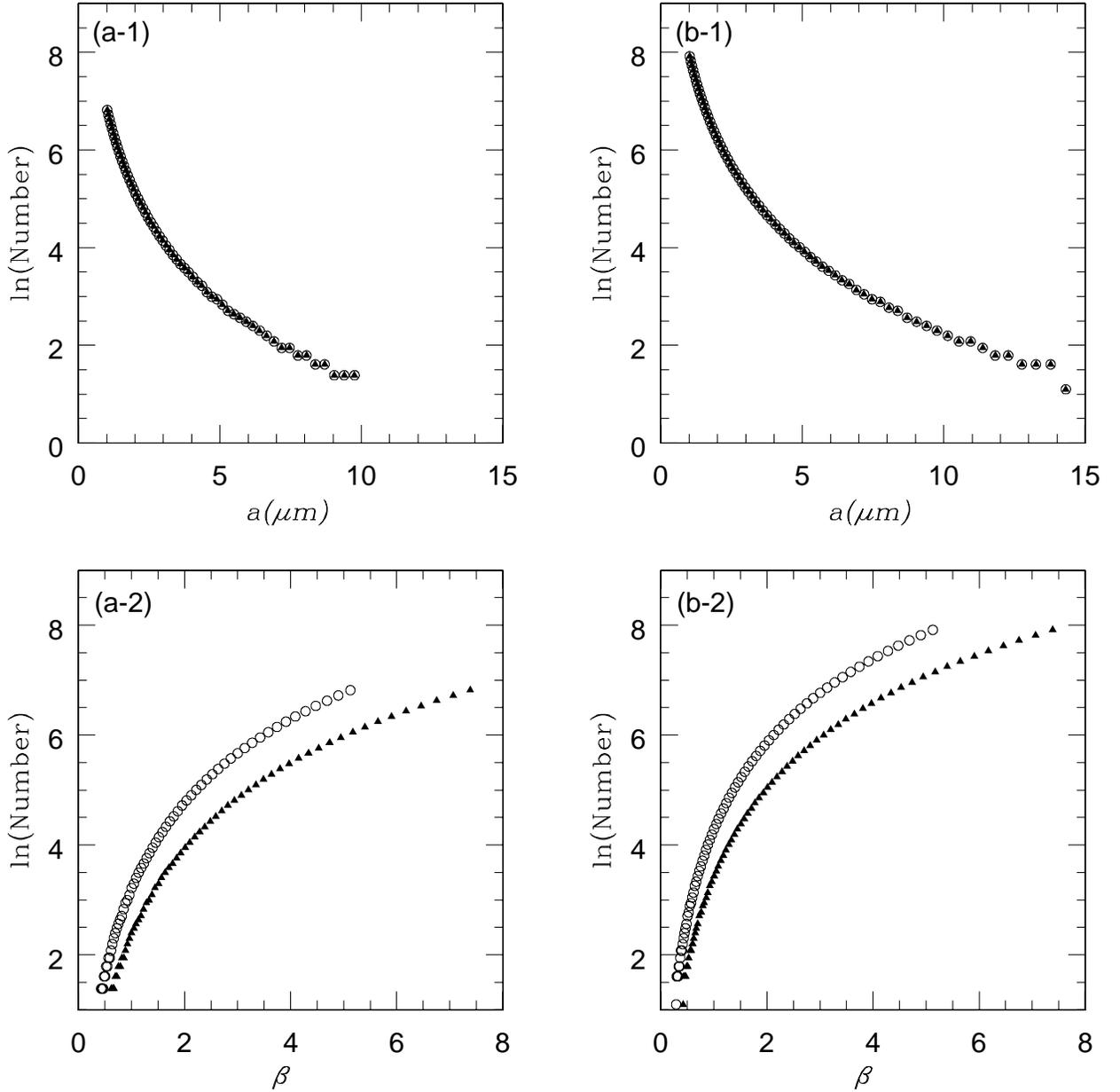,width=18.cm}
\caption[]{(a-1) The grain size distribution for the models with 
$a_{\rm max}=9.57 \mu m$; 
(a-2) The histograms of the values $\beta$ of models 
with $a_{\rm max}=9.57 \mu m$, where triangles 
are for those models with C400 and circles are for the models with
${\rm MgFeSiO_4}$;
(b-1) The grain size distribution for the models with 
$a_{\rm max}=14.04 \mu m$; 
(b-2) The histograms of the values $\beta$ of models 
with $a_{\rm max}=14.04 \mu m$, where triangles 
are for those models with C400 and circles are for the models with
${\rm MgFeSiO_4}$.
}
\end{figure}

\clearpage
\begin{figure}[htbp]
\psfig{figure=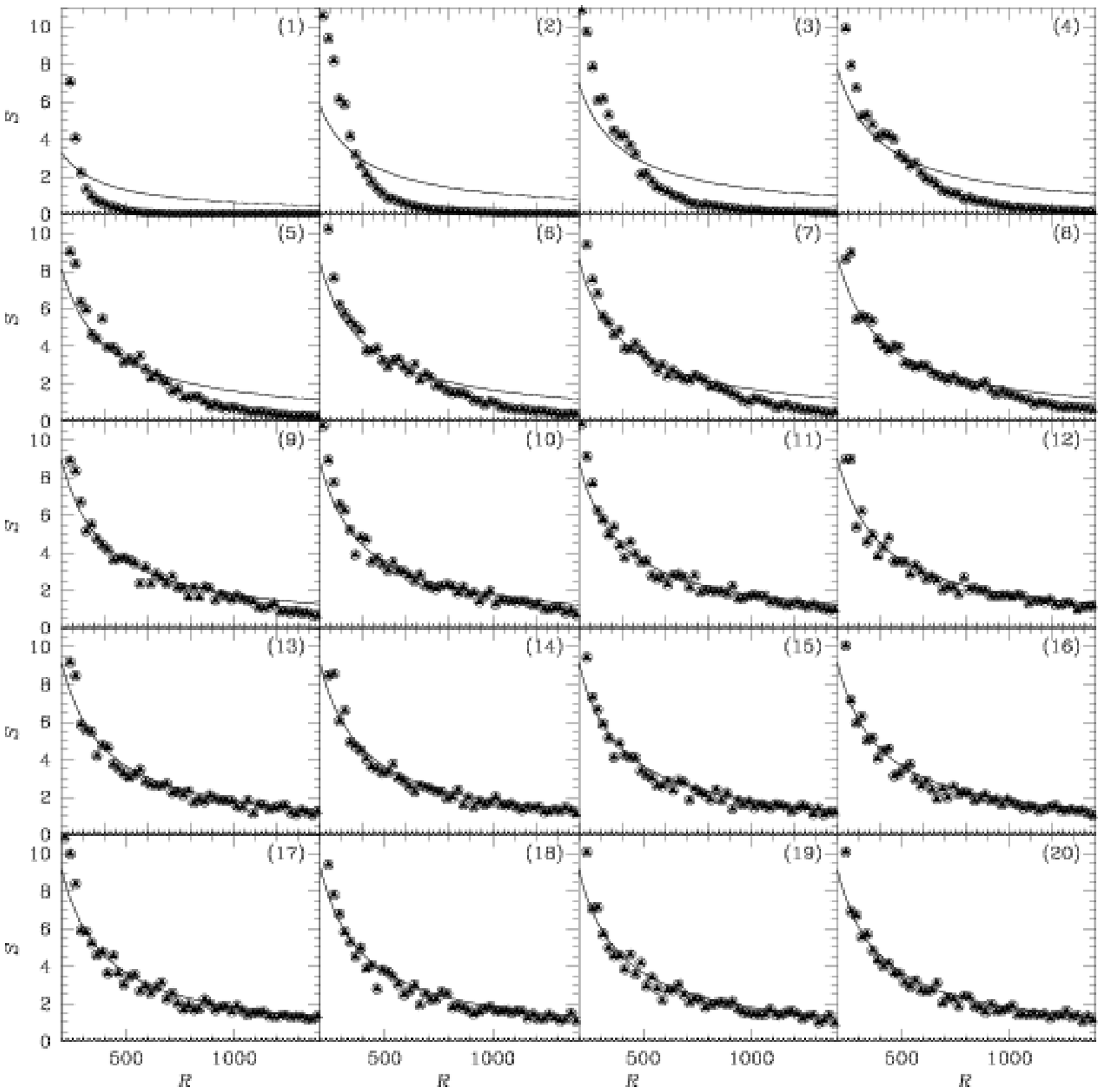,width=18.cm}
\caption[]{The surface mass densities of grains as functions of radii
of Model C2S.
The ith panel ($i=1,2,..., 20$) is at
the time $t=200\times i$, where
the solid curve is the best $1/R$ fitting function.
In all panels,
the triangles are for grains with
$\beta \geq 0.5$, the crosses are for grains with
$\beta < 0.5$, and the circles are the total. The unit of $R$ is AU and
the unit of $S$ is $10^{-12}{\rm g/AU^2}$. 
}  
\end{figure}
\clearpage
\begin{figure}[htbp]
\psfig{figure=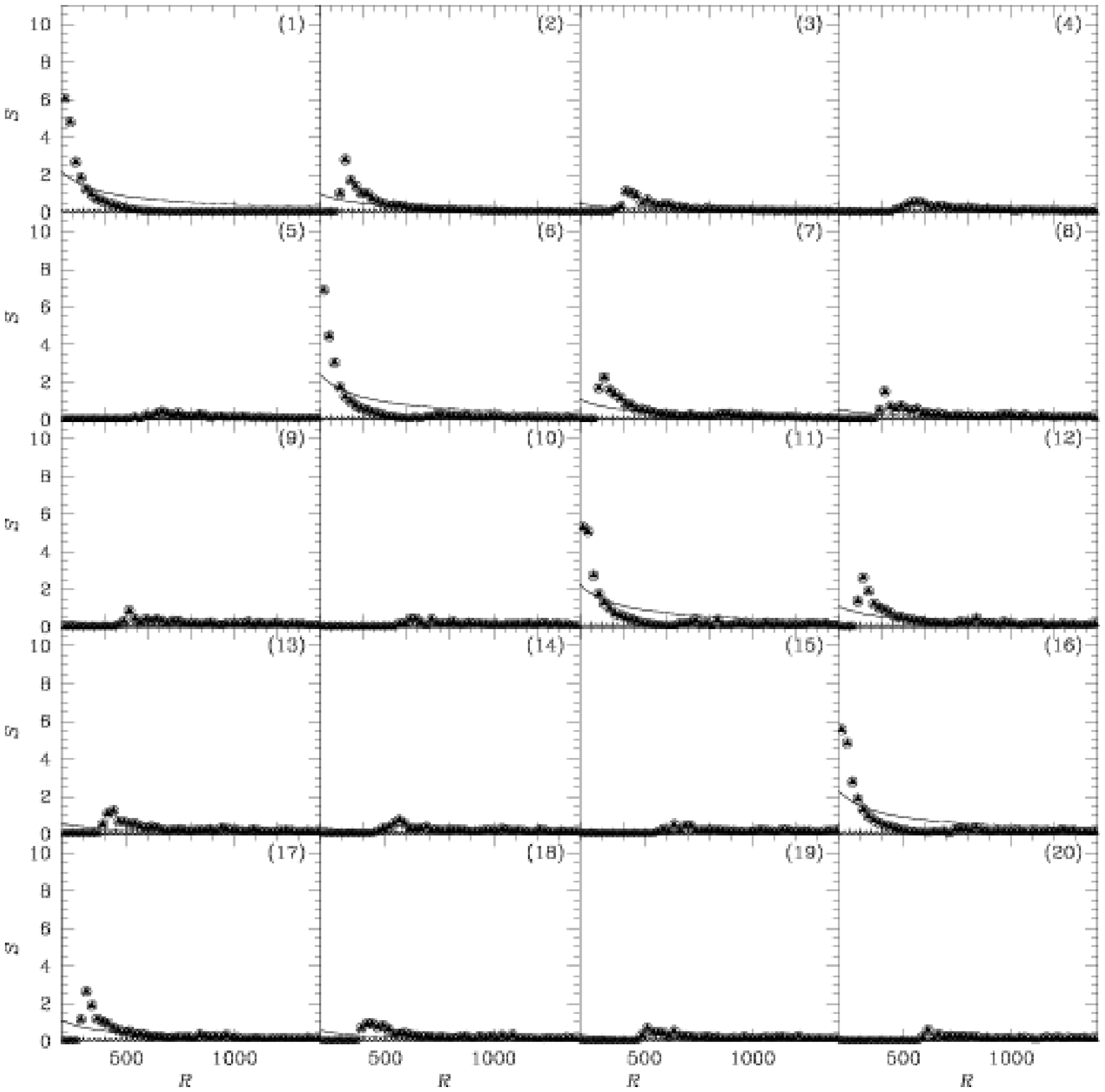,width=18.cm}
\caption[]{The surface mass densities of grains as functions of radii
of Model C3S.
The ith panel ($i=1,2,..., 20$) is at
the time $t=200\times i$, where
the solid curve is the best $1/R$ fitting function.
In all panels,
the triangles are for grains with
$\beta \geq 0.5$, the crosses are for grains with
$\beta < 0.5$, and the circles are the total. The unit of $R$ is AU and
the unit of $S$ is $10^{-12}{\rm g/AU^2}$. 
}                           
\end{figure}

\clearpage
\begin{figure}[htbp]
\psfig{figure=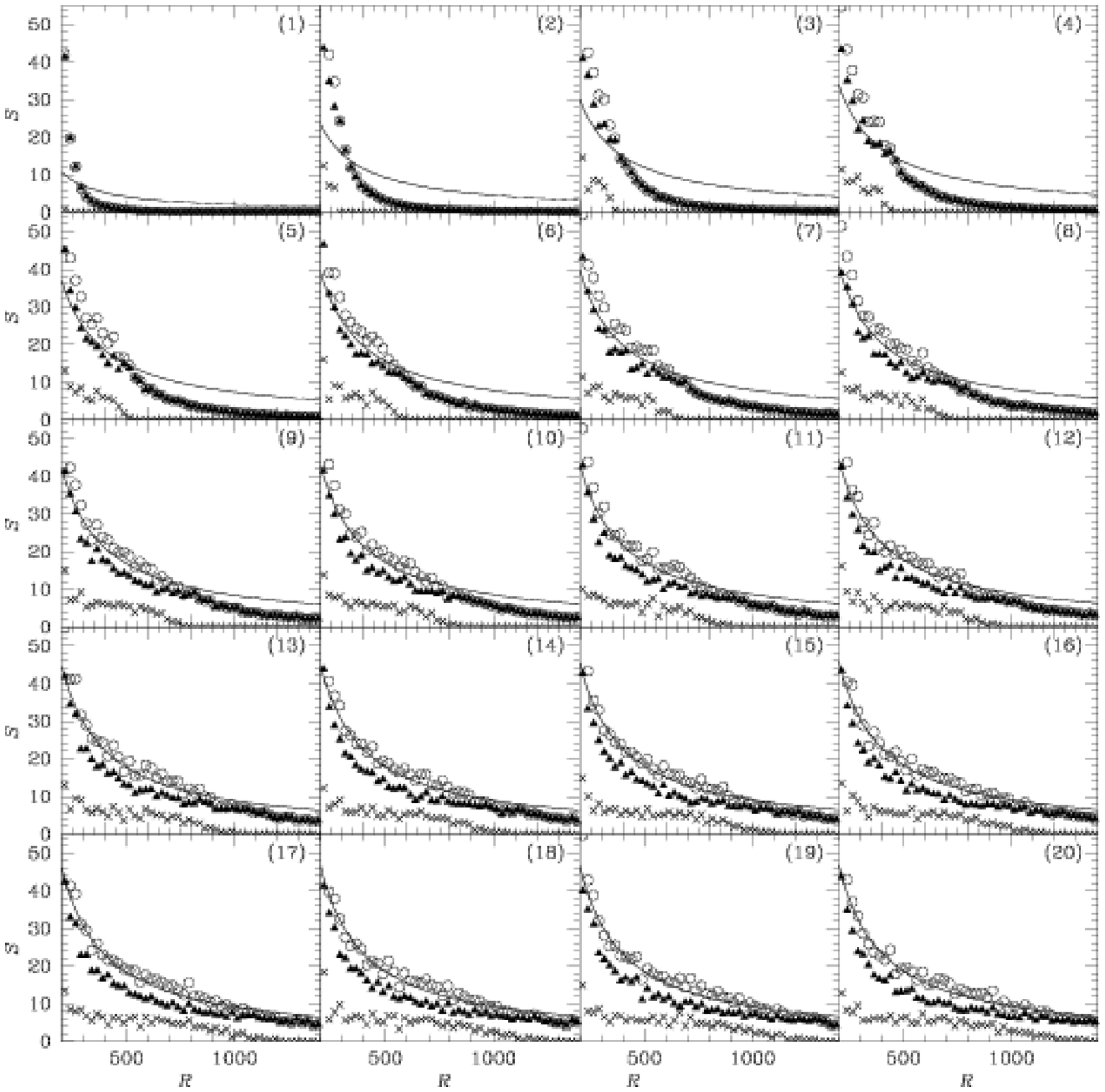,width=18.cm}
\caption[]{The surface mass densities of grains as functions of radii
of Model C2L.
The ith panel ($i=1,2,..., 20$) is at
the time $t=200\times i$, where
the solid curve is the best $1/R$ fitting function.
In all panels,
the triangles are for grains with
$\beta \geq 0.5$, the crosses are for grains with
$\beta < 0.5$, and the circles are the total. The unit of $R$ is AU and
the unit of $S$ is $10^{-12}{\rm g/AU^2}$. 
}                                                          
\end{figure}

\clearpage
\begin{figure}[htbp]
\psfig{figure=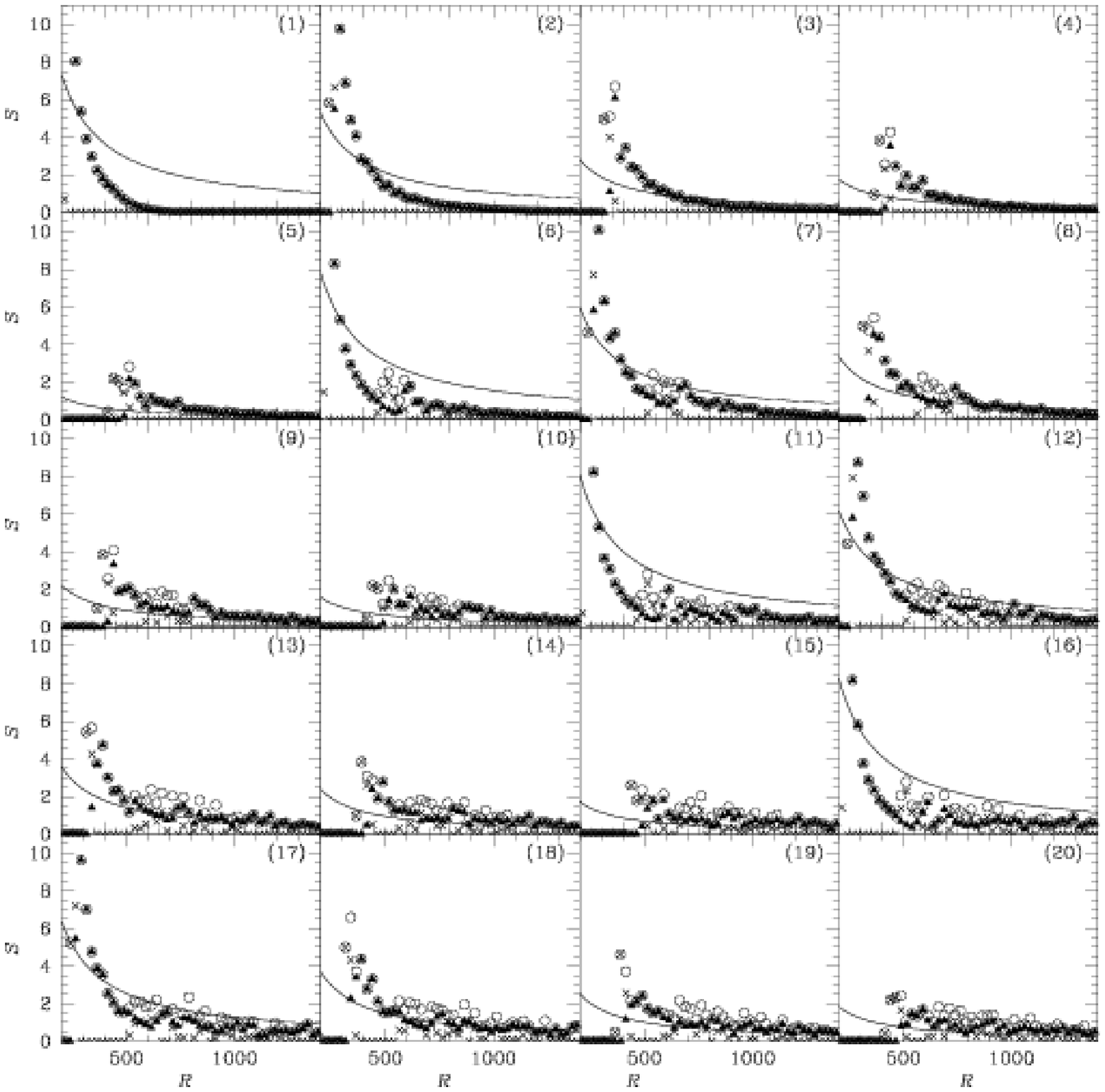,width=18.cm}
\caption[]{The surface mass densities of grains as functions of radii
of Model C3L.
The ith panel ($i=1,2,..., 20$) is at
the time $t=200\times i$, where
the solid curve is the best $1/R$ fitting function.
In all panels,
the triangles are for grains with
$\beta \geq 0.5$, the crosses are for grains with
$\beta < 0.5$, and the circles are the total. The unit of $R$ is AU and
the unit of $S$ is $10^{-12}{\rm g/AU^2}$. 
}                             
\end{figure}

\clearpage
\begin{figure}[htbp]
\psfig{figure=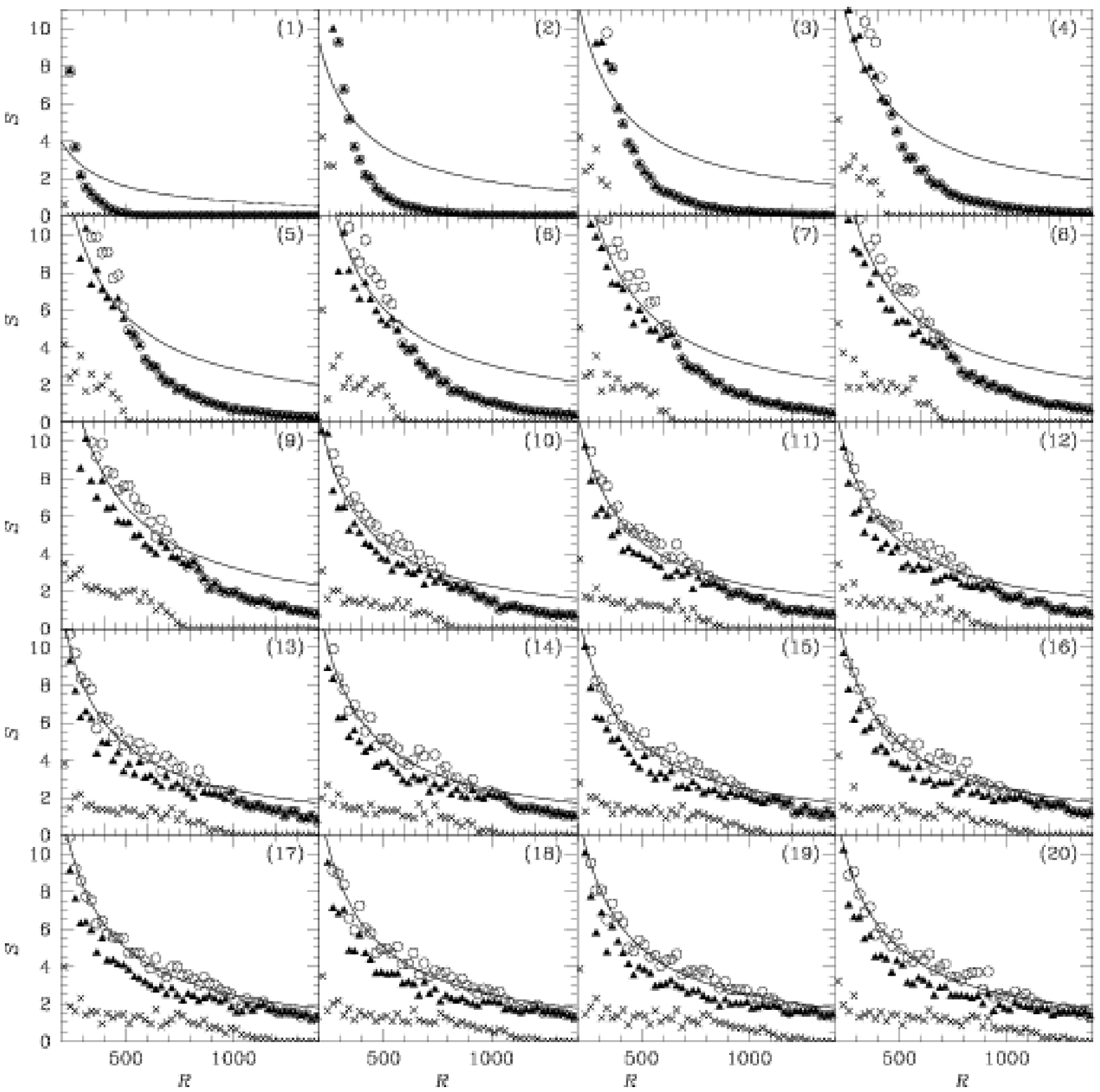,width=18.cm}
\caption[]{
The surface mass densities of grains as functions of radii
of Model Mg2S.
The ith panel ($i=1,2,..., 20$) is at
the time $t=200\times i$, where
the solid curve is the best $1/R$ fitting function.
In all panels,
the triangles are for grains with
$\beta \geq 0.5$, the crosses are for grains with
$\beta < 0.5$, and the circles are the total. The unit of $R$ is AU and
the unit of $S$ is $10^{-12}{\rm g/AU^2}$. 
}
\end{figure}

\clearpage
\begin{figure}[htbp]
\psfig{figure=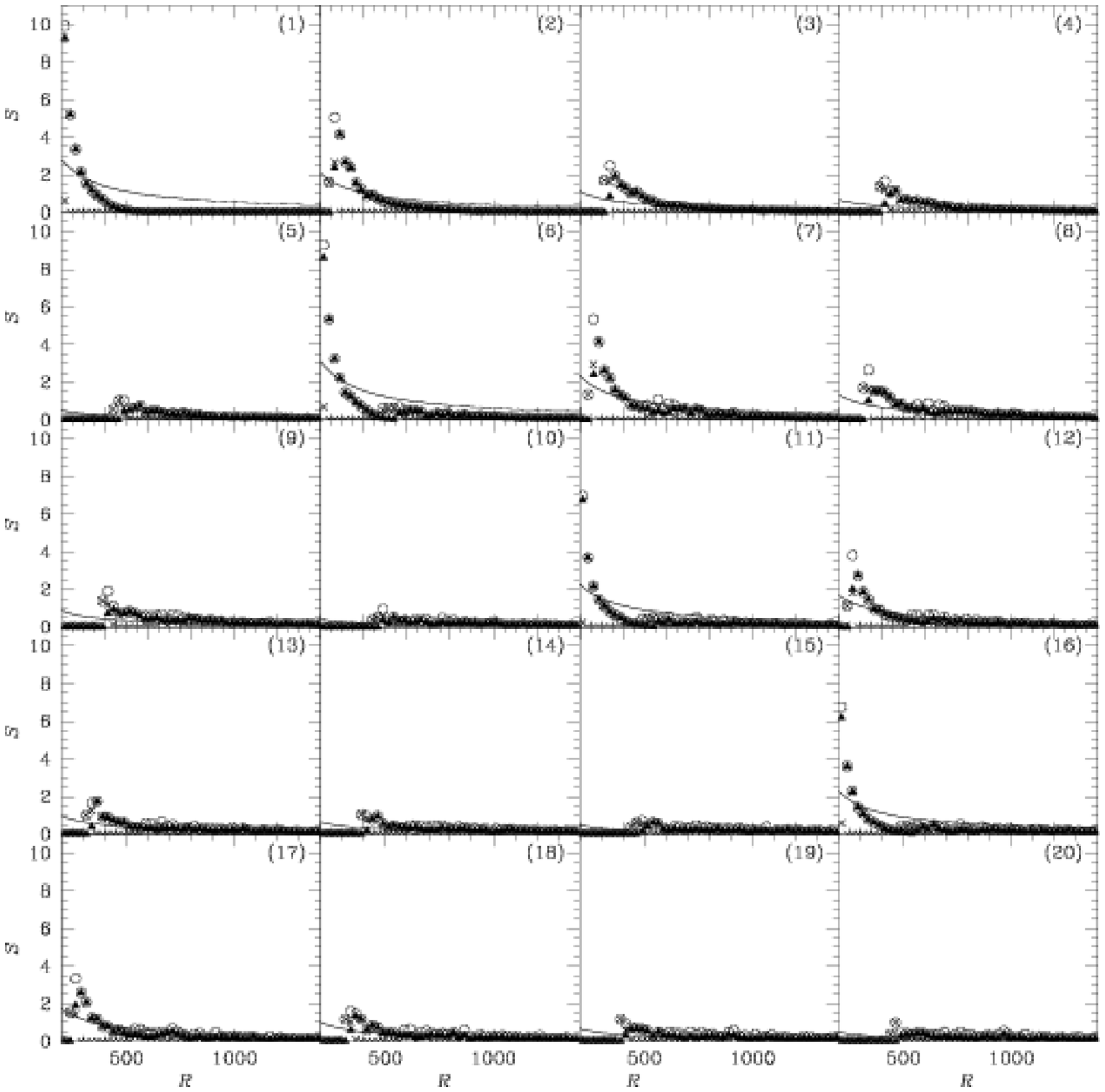,width=18.cm}
\caption[]{
The surface mass densities of grains as functions of radii
of Model Mg3S.
The ith panel ($i=1,2,..., 20$) is at
the time $t=200\times i$, where
the solid curve is the best $1/R$ fitting function.
In all panels,
the triangles are for grains with
$\beta \geq 0.5$, the crosses are for grains with
$\beta < 0.5$, and the circles are the total. The unit of $R$ is AU and
the unit of $S$ is $10^{-12}{\rm g/AU^2}$.  
}
\end{figure}

\clearpage
\begin{figure}[htbp]
\psfig{figure=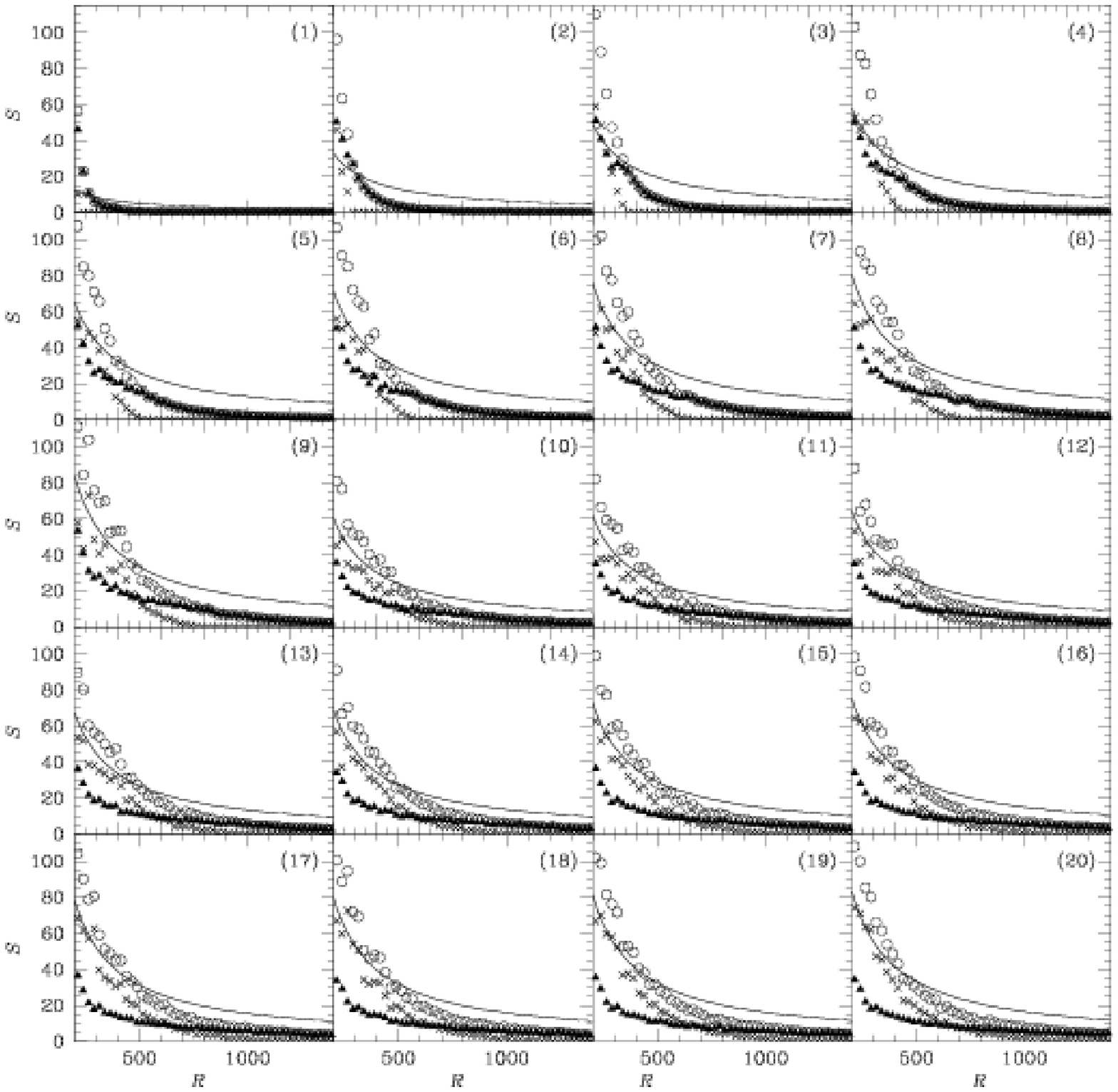,width=18.cm}
\caption[]{
The surface mass densities of grains as functions of radii
of Model Mg2L.
The ith panel ($i=1,2,..., 20$) is at
the time $t=200\times i$, where
the solid curve is the best $1/R$ fitting function.
In all panels,
the triangles are for grains with
$\beta \geq 0.5$, the crosses are for grains with
$\beta < 0.5$, and the circles are the total. The unit of $R$ is AU and
the unit of $S$ is $10^{-12}{\rm g/AU^2}$. 
}
\end{figure}

\clearpage
\begin{figure}[htbp]
\psfig{figure=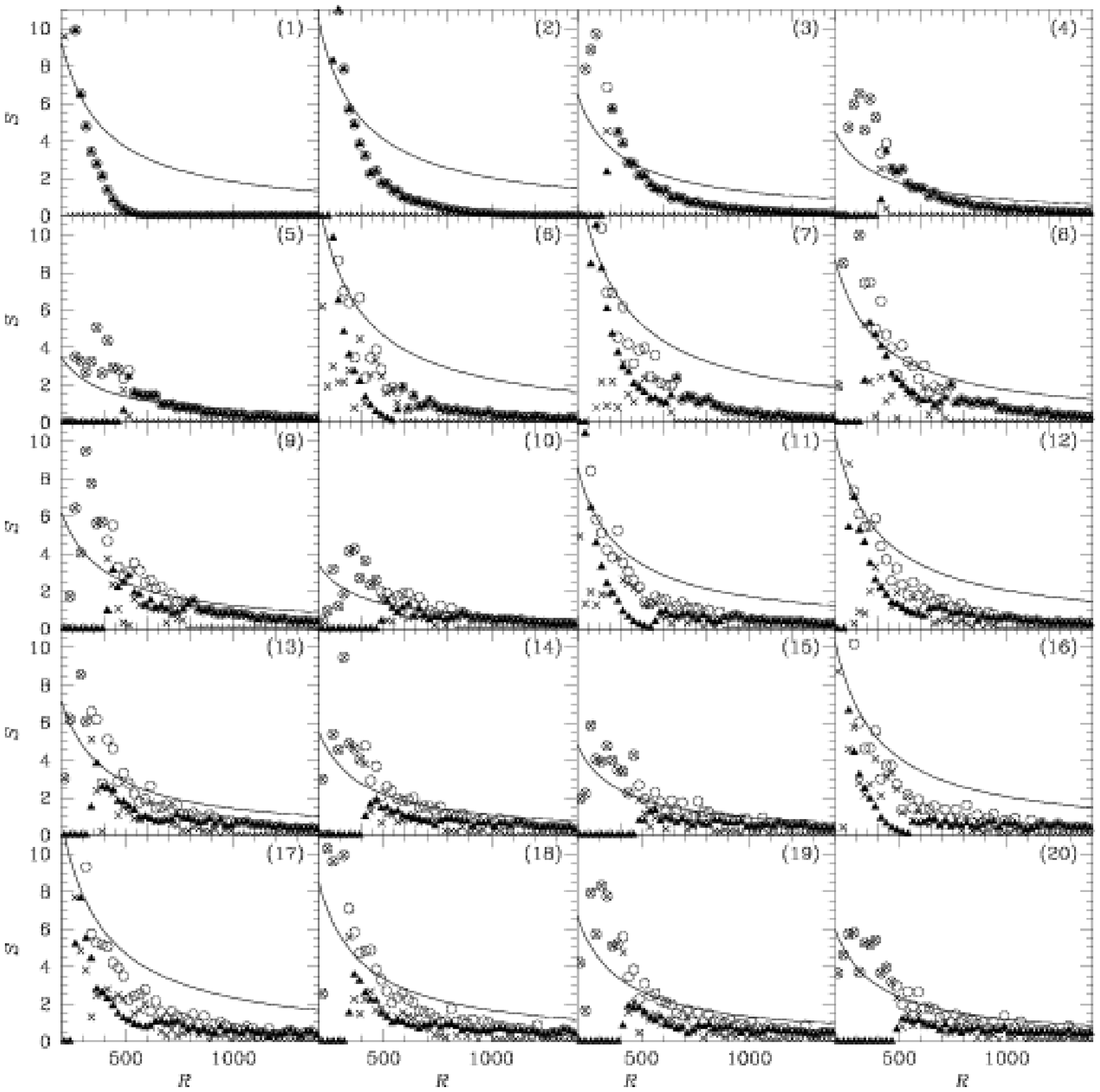,width=18.cm}
\caption[]{
The surface mass densities of grains as functions of radii
of Model Mg3L.
The ith panel ($i=1,2,..., 20$) is at
the time $t=200\times i$, where
the solid curve is the best $1/R$ fitting function.
In all panels,
the triangles are for grains with
$\beta \geq 0.5$, the crosses are for grains with
$\beta < 0.5$, and the circles are the total. The unit of $R$ is AU and
the unit of $S$ is $10^{-12}{\rm g/AU^2}$. 
}
\end{figure}

\clearpage
\begin{figure}[htbp]
\psfig{figure=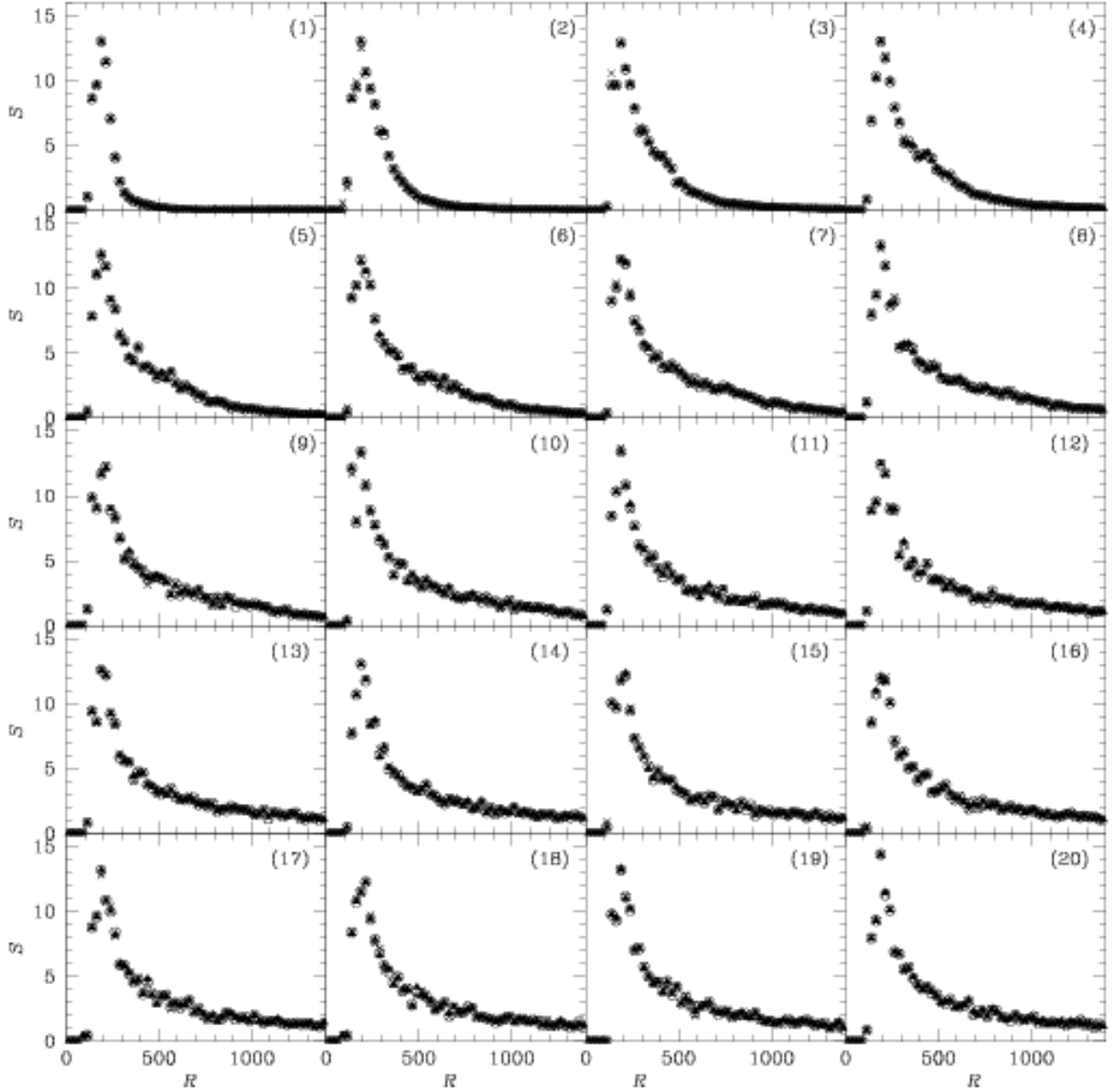,width=18.cm}
\caption[]{
The surface mass densities of grains as functions of radii
of Run 1, 2, and Model C2S.
The ith panel ($i=1,2,..., 20$) is at
the time $t=200\times i$.
In all panels, the crosses are for Run 1,
the triangles are for Run 2, and
the circles are for Model C2S. The unit of $R$ is AU and
the unit of $S$ is $10^{-12}{\rm g/AU^2}$. 
}
\end{figure}

\clearpage
\begin{figure}[htbp]
\psfig{figure=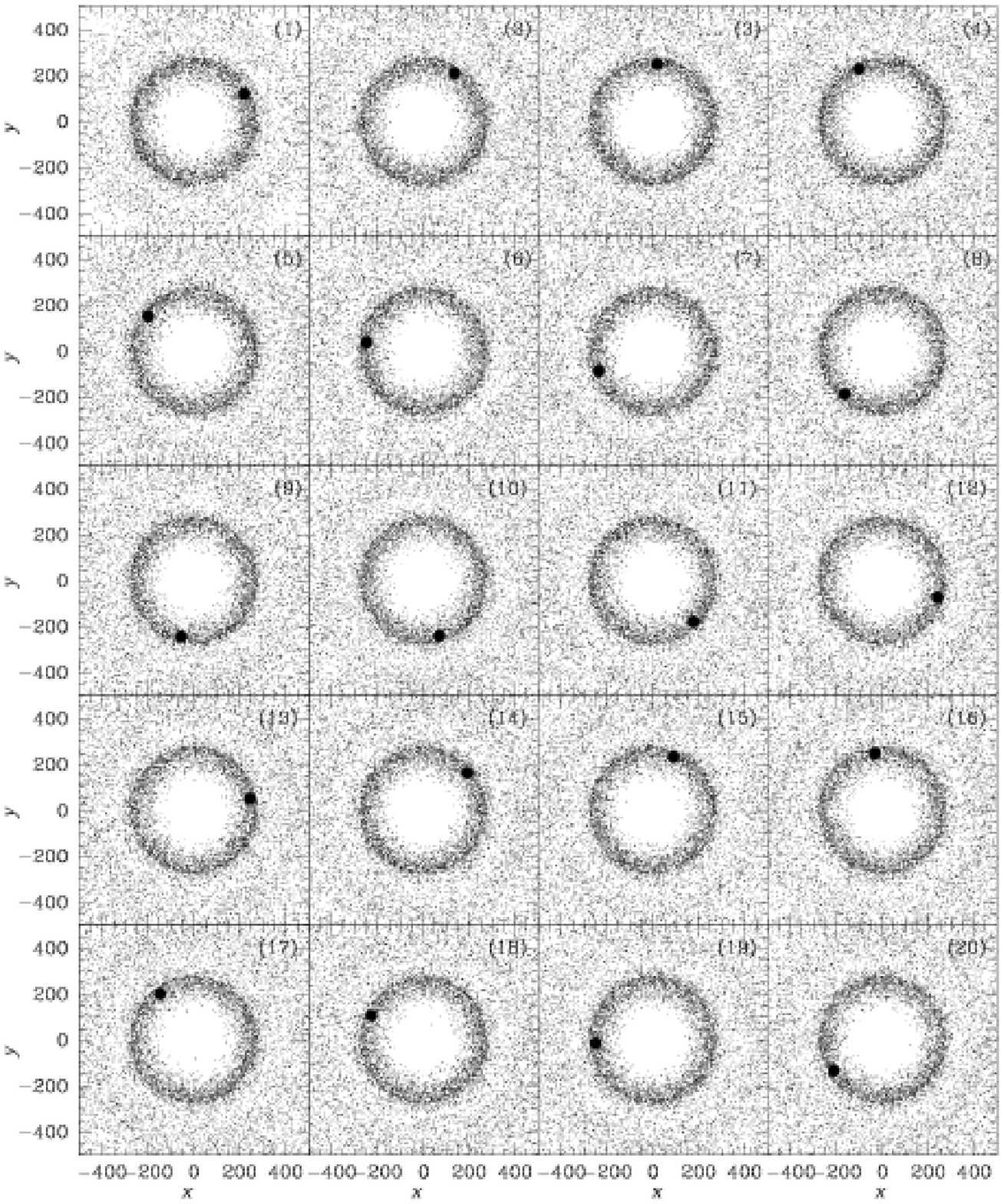,width=21.cm}
\caption[]{The grain distributions on the $x-y$ plane of Run 2.
The ith panel ($i=1,2,..., 20$) is at
the time $t=200\times i$.
In all panels, the dots show the locations of grains, 
and the full circle represents the planet.
The unit of both axes is AU.
}
\end{figure}

\clearpage
\begin{figure}[htbp]
\psfig{figure=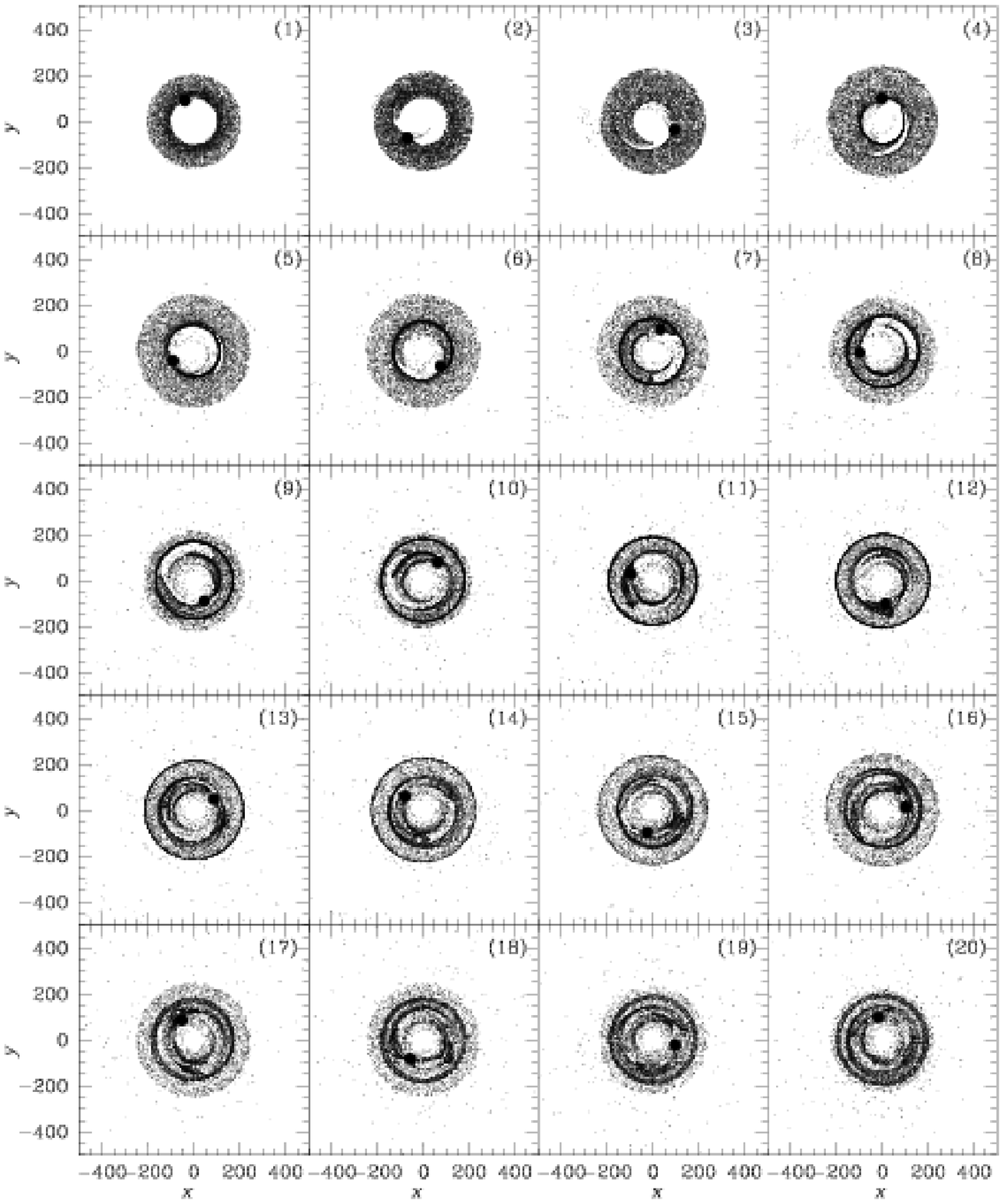,width=21.cm}
\caption[]{The grain distributions on the $x-y$ plane of Run 3.
The ith panel ($i=1,2,..., 20$) is at the time $t=200\times i$.
In all panels, the dots show the locations of grains, 
and the full circle represents the planet.
The unit of both axes is AU.
}
\end{figure}

\clearpage
\begin{figure}[htbp]
\psfig{figure=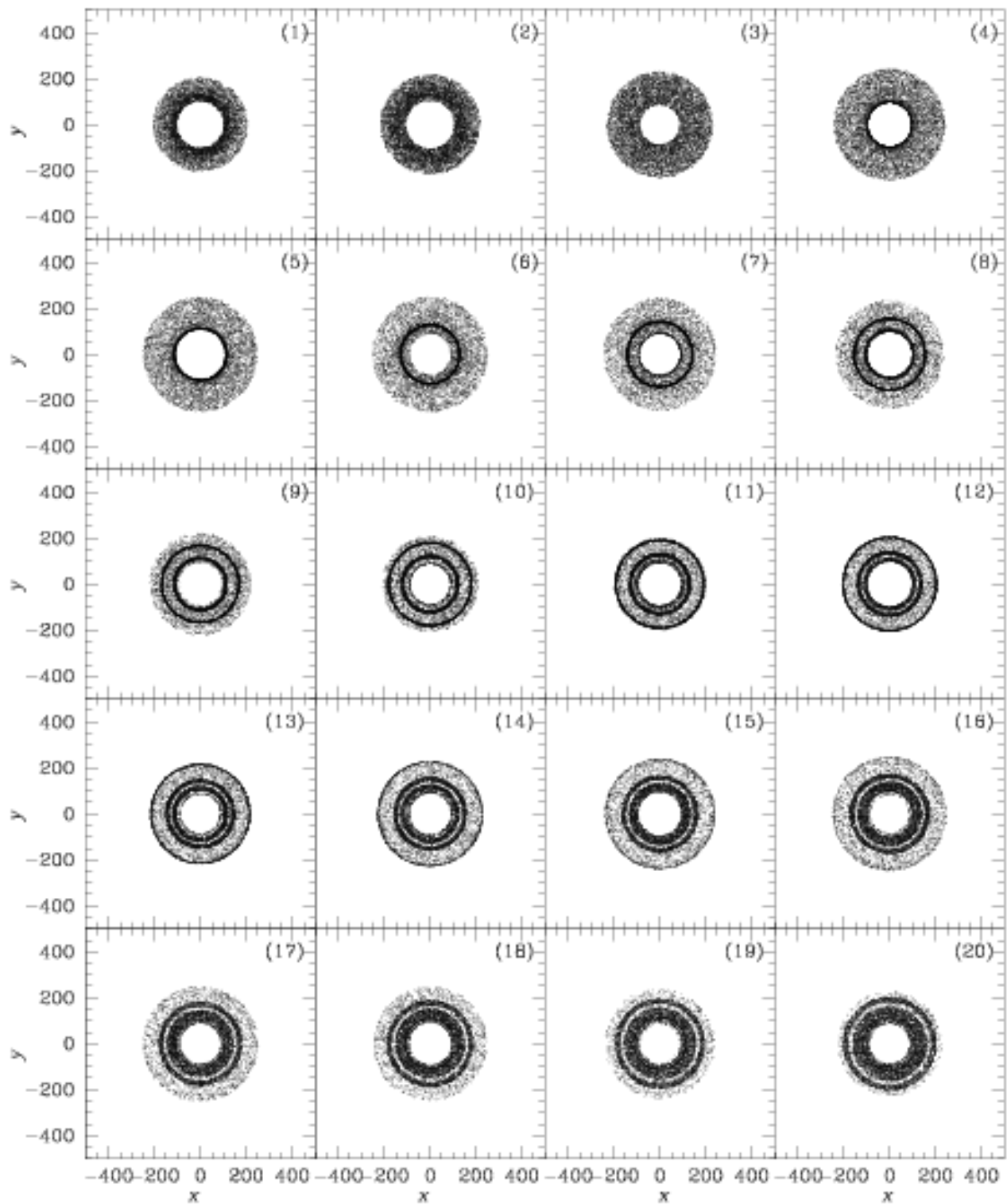,width=21.cm}
\caption[]{The grain distributions on the $x-y$ plane of Run 4.
The ith panel ($i=1,2,..., 20$) is at the time $t=200\times i$.
In all panels, the dots show the locations of grains.
The unit of both axes is AU.
}
\end{figure}

\clearpage
\begin{figure}[htbp]
\psfig{figure=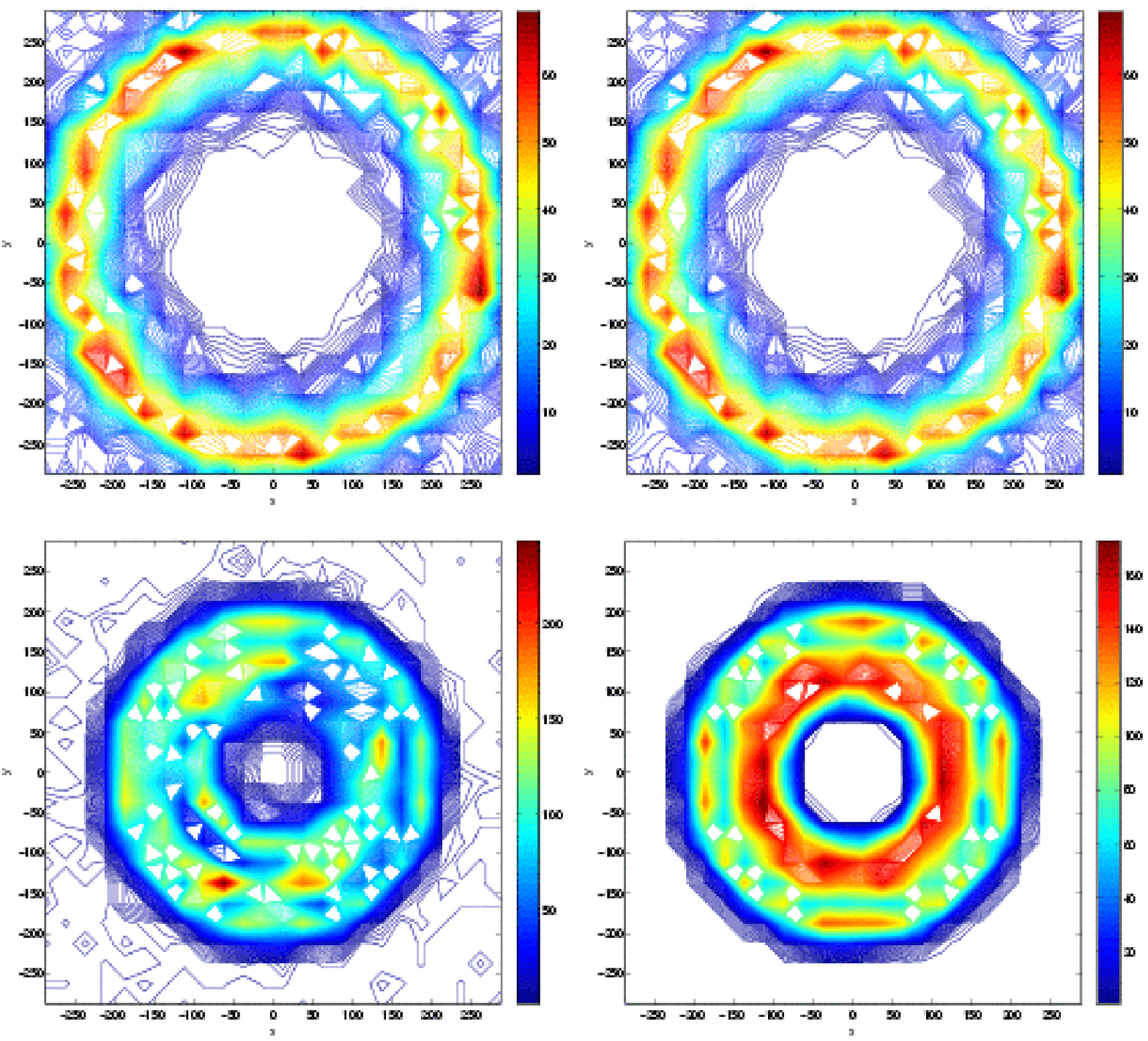,width=18.cm}
\caption[]{
The color contour of the grain particle distribution at $t=4000$,
where the top-left, top-right, bottom-left, 
and bottom-right panel is for Model C2S, Run 2, Run 3, and Run 4,
respectively.
}
\end{figure}


\clearpage

\begin{thebibliography}{21}

\bibitem{Ar1}
Artymowicz, P. 1997, Annual Review of Earth and Planetary Sciences, 
25, 175 

\bibitem{Ar2}
Artymowicz, P., Clampin, M. 1997, ApJ, 490, 863 


\bibitem{A1} Aumann, H. H. et al. 1984, ApJ, 278, L23

\bibitem{B} Burns, J. A., Lamy, P. L., Soter, S. 1979, Icarus, 40, 1

\bibitem{C} Cheney, W., Kincaid, D. 1998, 
Numerical Mathematics and Computing, 4th Edition,
International Thomson Publishing  

\bibitem{C1} Cuzzi, J. N., Dobrovolskis, A. R., Champney, J. M.
1993, Icarus, 106, 102 
               
\bibitem{G1} Gozdziewski, K., Maciejewski, A. J. 2001, ApJ, 563, L81

\bibitem{G2} Greaves, J. S., Holland, W. S., Moriarty-Schieven, G., 
Jenness, T., Dent, W. R. F., Zuckerman, B., McCarthy, C., Webb, R. A., 
Butner, H. M., Gear, W. K., Walker, H. J.  1998, ApJ, 506, L133
 
\bibitem{G3}
Grigorieva, A., Artymowicz, P., Thebault, P. 2007,
A\&A, 461, 537


\bibitem{H} 
Harvey, P. M., Wilking, B. A., Joy, M. 1984, \nat, 307, 441 

\bibitem{H1} Hatzes, A. P., Cochran, W. D., McArthur, B., Baliunas, S. L., 
Walker, G. A. H., Campbell, B., Irwin, A. W., Yang, S., Kurster, M., 
Endl, M., Els, S., Butler, R. P., Marcy, G. W. 2000, ApJ, 544, L145

\bibitem{H2} 
Heinrichsen, I., Walker, H. J., Klaas, U. 1998, \mnras, 293, L78 

\bibitem{H3} 
Holland, W. S. et al. 1998, \nat, 392, 788 


\bibitem{I1} Ishimaru, A. 1991, 
Electromagnetic Wave Propagation, Radiation, and Scattering,
London: Prentice-Hall International, Inc. 

\bibitem{Ji} Ji, J., Kinoshita, H., Liu, L., Li, G. 2007, 
ApJ, 657, 1092

\bibitem{Ji1} Ji, J., Li, G., Liu, L. 2002, ApJ, 572, 1041


\bibitem{Jiang2003}
Jiang, I.-G., Ip, W.-H., Yeh, L.-C. 2003, ApJ, 582, 449

\bibitem{JiangY1}  Jiang, I.-G., Yeh, L.-C. 2004a, AJ, 128, 923

\bibitem{JiangY2}  Jiang, I.-G., Yeh, L.-C. 2004b,
Int. J. Bifurcation and Chaos, 14, 3153

\bibitem{JiangY3}  Jiang, I.-G., Yeh, L.-C. 2004c,
MNRAS, 355, L29

\bibitem{K} Kinoshita, H., Nakai, H. 2001, PASJ, 53, L25

\bibitem{K1} 
Koerner, D. W., Sargent, A. I., Ostroff, N. A. 2001, \apjl, 560, L181 

 
\bibitem{L} Laor, A., Draine, B. T. 1993, ApJ, 402, 441

\bibitem{L1} Laughlin, G., Chambers, J. 2001, ApJ, 551, L109


\bibitem{M} 
Mauron, N., Dole, H. 1998, \aap, 337, 808 

\bibitem{Amaya2002} 
Moro-Martin, A., Malhotra, R. 2002, AJ, 124, 2305 

\bibitem{Amaya2005} 
Moro-Martin, A., Wolf, S., Malhotra, R. 2005,
ApJ, 621, 1079 




\bibitem{S1} Su, K. Y. L. et al. 2005, ApJ, 628, 487


\bibitem{T1} Takeuchi, T., Artymowicz, P. 2001, ApJ, 557, 990

\bibitem{T2} Takeuchi, T., Lin, D. N. C. 2002, ApJ, 
581, 1344
	
	
\bibitem{T3} Thebault, P., Augereau, J.-C. 2007, A\&A, 472, 169

\bibitem{V1} Van de Hulst, H. C. 1957, 
Light Scattering by Small Particles, New York: Wiley

\bibitem{V2} 
Van der Bliek, N. S., Prusti, T., Waters, L. B. F. M. 1994, 
\aap, 285, 229 

\bibitem{W1} 
Wilner, D. J., Holman, M. J.,
Kuchner, M., Ho, P. T. P. 2002, \apjl, 569, L115 


 


	


\bibitem{Z} 
Zakamska, N. L., Tremaine, S. 2004, AJ, 128, 869

\bibitem{Z1} 
Zuckerman, B., Becklin, E. E. 1993, \apj, 414, 793 

\end{thebibliography}
\end{document}